\documentclass[eqsecnum,twocolumn,aps,tightenlines,floats]{revtex4}
\def\bea{\begin{eqnarray}}
\def\eea{\end{eqnarray}}
\usepackage{graphics}
\usepackage{graphicx}
\usepackage{epsf} 
\begin{document} 

\phantom{0}
\vspace{-0.5in}
\hspace{5in}\parbox{1.5in}{ \leftline{JLAB-THY-03-205}
                \leftline{WM-03-112}
\vspace{-1in}}
\title
{\bf Electromagnetic interactions of three-body systems in the
covariant spectator theory}

\author{Franz Gross$^{1,2}$, Alfred
Stadler$^{3,4}$, and M.~T.~Pe\~na$^{5}$  }

\address{
$^1$College of William and Mary, Williamsburg, 
Virginia 23185
\\
$^2$Thomas Jefferson National Accelerator 
Facility, Newport News, VA 23606
\\
$^3$Centro de F\'\i sica Nuclear da Universidade 
de Lisboa, 1649-003 Lisboa, Portugal
\\
$^4$Departamento de F\'\i sica da Universidade 
de \'Evora, 7000-671 \'Evora, Portugal
\\
$^5$Departamento de F\'\i sica e CFIF, 
Instituto Superior T\'ecnico, 1049-001 Lisboa, Portugal
}

\date{\today}

\begin{abstract} 

We derive a complete Feynman diagram expansion for the elastic
form factor and the three-body  photo and
electrodisintegration of the three-body bound state using the
covariant spectator theory.  We show that the equations obtained
are fully consistent with bound-state equations and the
normalization condition previously derived for the covariant 
three-body bound state, and that the results conserve current.    

\end{abstract}
 
 
\phantom{0}
\vspace{7.0in}
\vspace{-6in}
\maketitle

\section{Introduction} 

The covariant spectator formalism \cite{Gr69} has been applied
successfully to the description of $NN$ scattering and the
deuteron bound state \cite{BG79,GVOH}, the deuteron form factors
\cite{ACG80,VODG95} electrodisintegration of the deuteron
\cite{AGJ02}, and the three-body system
\cite{Gr82}, with emphasis on the numerical solution
of the three-body bound state equations \cite{SG97,SGF97}. 
While most of the theory has been developed using diagrammatic
methods, all of the two-body theory \cite{AVOG98} and the 
covariant normalization condition for three-body
spectator wave functions \cite{AGS97} have also been
derived algebraicially.  Kvinikhidze and Blankleider have also
studied the current for both the two and three-body spectator
equations \cite{KB97a,KB97} using the gauging of
equations method. 

In this paper we use the same diagrammatic method that is the basis of the results
given in Refs.~\cite{Gr69,BG79,GVOH,ACG80,VODG95,AGJ02,Gr82,SG97,SGF97}  to obtain
the complete set of gauge invariant Feynman diagrams needed to evaluate the
three-body form factors and three-body breakup amplitudes for three-body photo or
electrodisintegration.  Our results agree with those recently dervied algebraically
by Adam and Van Orden \cite{AVO03}, and, although they have a different form, can
also be shown to agree with the results  of Kvinikhidze and Blankleider
\cite{KB97}.   These results, together with the relativistic wave functions solutions
found in Ref.~\cite{SG97}, are currently being used \cite{GPS} to study the three
body electrodisintegration process $^3$He$(e,e'pp)n$ recently measured at the
Jefferson Laboratory \cite{Wein03}. 

This paper is divided into five sections and a long Appendix.  Following this short
introduction,  we first review the definitions and some properties of the covariant
three-body spectator subamplitudes and spectator equations.  In Sec.\ III, the basic
results for both the elastic and inelastic three-body currents are derived from an
analysis of the infinite series of Feynman diagrams that consistently defines both
the scattering equation and the currents.   The derivation is based on the gauging
method of  Kvinikhidze and Blankleider \cite{KB97} (referred to as KB throughout this
paper), with a detailed diagrammatic proof of gauge invariance and a full discussion
of double counting.  Then, in Sec.\ IV, explicit algebraic forms of the final results
for the currents are presented and in Sec.\ V conclusions, and a detailed comparison
with KB  are given.  An algebraic proof that the elastic current conserves the charge
of the three-body bound state is given in the Appendix.

\section{working with spectator three-body amplitudes}

\subsection{Covariant Faddeev subamplitudes}

In this paper $\{i,j,k\}$ denote any 
permutation of particles $1,2,3$, so that $j\ne i$,
$k\ne i$ and $j\ne k$, and, for example,
$i$ can represent any of the three particles. Then, in
the nonrelativistic Faddeev theory, the full three-body
vertex function, which we denote
$\left|\Gamma\right>$, is  decomposed into three
subamplitudes
$\left|\Gamma^i\right>$ which denote that part of the
vertex in which particle $i$ is the {\it spectator\/} and
the {\it other\/} two particles ($j$ and $k$) were the last
to interact.  The full vertex is a sum of the three
subvertices
\begin{eqnarray}
\left|\Gamma\right>=\sum_i\left|\Gamma^i\right>=
\left|\Gamma^1\right> + \left|\Gamma^2\right> +
\left|\Gamma^3\right>\, .
\end{eqnarray}

\begin{figure}
\leftline{
\mbox{
\includegraphics[width=3in]{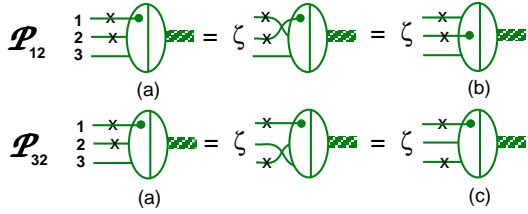}
}}
\vspace*{-0.2in}
\caption{\footnotesize\baselineskip=10pt (Color online) Figure (a) shows
the notation for the vertex subamplitude $\Gamma_2^1$.  Lines
marked by an $\times$ are on mass-shell, and the spectator
particle is the one that connects to the small dot inside the
oval.  Note that
${\cal P}_{12}\Gamma_2^1=\zeta\Gamma_1^2$, as shown in (b), and
${\cal P}_{32}\Gamma_2^1=\zeta\Gamma_3^1$, as shown in (c).}
\label{amps}
\end{figure}

In the covariant spectator theory \cite{Gr82,SG97}, the
spectator is on-shell, and {\it one of the two interacting
particles is also on-shell\/}.  Hence there are now 6
possible Faddeev subvertex functions, denoted by
$\left|\Gamma^i_j\right>$, where $i$ is the (on-shell)
spectator and $j$ is also on-shell, so that only one of
the three particles, $k$ in this example, is off-shell.  The
diagrammatic representation of this amplitude is shown in
Fig.~\ref{amps}(a).  If the three particles are identical (as we
assume in this paper) it can be shown
\cite{SGF97}, under the permutation operator ${\cal P}_{ij}$ that
interchanges particle $i$ and $j$, that 
\begin{eqnarray}
{\cal P}_{ij}\left|\Gamma^i_j\right>=&& \zeta
\left|\Gamma^j_i\right>\nonumber\\
{\cal P}_{jk}\left|\Gamma^i_j\right>=&& \zeta
\left|\Gamma^i_k\right> \nonumber\\
{\cal P}_{ik}\left|\Gamma^i_j\right>=&& \zeta
\left|\Gamma^k_j\right> \label{permsymm}
\end{eqnarray}
where $\zeta=\pm$ depending on whether or not the particles are bosons or fermions [a
diagrammatic representation of the first two of these equations is given in
Fig.~\ref{amps}].  In practice, this means that the particles may be freely relabeled
in Feynman diagrams (as we will discuss below). 

\begin{figure}
\begin{center}
\mbox{
\includegraphics[width=2in]{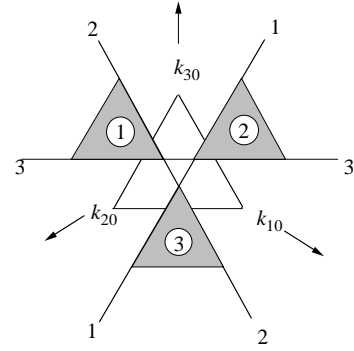}
}
\end{center}
\caption{\footnotesize\baselineskip=10pt The three shaded areas
are kinematically allowed regions of the energies of the three
particles, subject to the constraint that their total is $M_B<3m$,
and that two are on-shell. 
[To read the figure, note that each point on the plane
defines three unique energies. The energy of each particle
is the perpendicular distance from one of the sides of the
large central triangle (positive if ``above'' the side and
negative if ``below'', where the positive direction is
shown by the arrows).  The geometry of this Dalitz-like
plot insures that the sum of the three energies at each
point equals the altitude of the triangle (chosen to be
$M_B$).]  The on-shell condition $k_{i0}>m$ requires that
$k_{i0}$ lie beyond the line labeled with the number $i$
at each end, and the shaded areas labeled by the number
$i$ (equal to 1, 2, or 3) are regions where particle $i$
is off-shell, and $j$ and $k$ have energies $E>m$.  Note
that these three areas do not overlap.}
\label{energydom}
\end{figure}

The subvertex functions in the spectator theory each
satisfy a different constraint, and care must be taken to
match this constraint to the physics.  For example, the
energy of particle 3 in the subvertex function
$\left|\Gamma^1_2\right>$ is $k_{30}=M_B-E_1-E_2$, where
throughout this paper $E_i=\sqrt{m^2+{\bf k}_i^2}$ will
always represent a physical energy, $M_B$ is the bound
state mass, and ${\bf k}_i$ are the three-momenta of the
three particles in the three-body c.m.  However, the
energy of particle 3 in the subvertex functions
$\left|\Gamma^1_3\right>$ and $\left|\Gamma^3_1\right>$
is $E_3$.  The energy domains of each of the
subamplitudes is illustrated diagrammatically in Fig.\
\ref{energydom}.  Note that, if $M_B<3m$, there are three distinct
domains that do not overlap.  

We now describe briefly how to derive the bound state and
scattering equations for three identical particles.  To make the
discussion simple and intuitive, we use a diagrammatic approach
(similar to that used in the original Ref.\ \cite{Gr82}). 

\subsection{The three-body scattering equation}
\label{IIB}

\begin{figure}
\leftline{
\mbox{
\includegraphics[width=3.in]{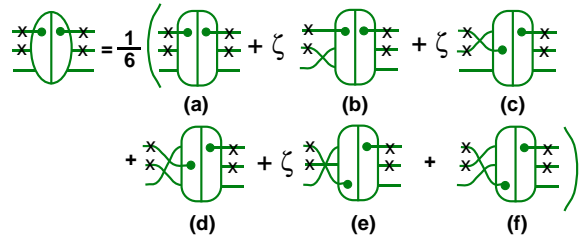}
}}
\caption{\footnotesize\baselineskip=10pt (Color online) 
Diagrammatic representation of the construction of the {\it
symmetrized\/} three-body scattering
subamplitude ${T}^{11}_{22}$ (represented by the oval) from
{\it unsymmetrized\/} subamplitudes ${\cal T}^{ii'}_{jj'}$
(represented by rounded rectangles).  Only the symmetrization
of the final state is shown.  The 6 figures correspond to the
6 terms in Eq.~(\ref{A3}) as follows: (a) $\to$ 1, (b)
$\to\zeta{\cal P}_{23}$, (c) $\to\zeta{\cal P}_{12}$, 
(d) $\to{\cal P}_{23}{\cal P}_{12}$, 
(e) $\to\zeta{\cal P}_{13}$, and (d) $\to{\cal P}_{12}{\cal
P}_{23}$.}
\label{fig:symm}
\end{figure}  

For simplicity, this paper addresses only cases in which the 
three-body scattering is a succession of two-body
scatterings.  This means that three-body forces of relativistic origin, 
as defined in
Ref.~\cite{Gr82}, will be neglected. 
 
We begin with a review of some of the results of Ref.~\cite{AGS97}.  The quantity of
primary interest is the {\it symmetrized\/} three-body subamplitude ${
T}^{ii'}_{jj'}$, which can be obtained from the {\it unsymmetrized\/} subamplitudes
${\cal T}^{ii'}_{jj'}$.  Here the superscripts $i',i$  denote the on-shell spectators
in the initial and final state, respectively, and the subcripts $j',j$ the on-shell
interacting particle in the initial and final state, respectively.  The {\it
symmetrized\/} subamplitude can be obtained from the {\it unsymmetrized\/}
subamplitude by action of the three-particle antisymmetrization projection operator

\begin{eqnarray}
{\cal A}_3=\frac{1}{6}\Big\{1 +\zeta{\cal P}_{12} +\zeta{\cal
P}_{13} + \zeta{\cal P}_{23} \nonumber\\
+ {\cal P}_{12}{\cal
P}_{23} + {\cal P}_{23}{\cal P}_{12}\Big\}\,  \label{A3}
\end{eqnarray} 
normalized to $({\cal A}_3)^2={\cal A}_3$.  The fully symmetrized amplitude is
\begin{eqnarray}
T^{ii'}_{jj'}={\cal A}_3 \,{\cal T}^{ii'}_{jj'}\, {\cal A}_3 \, . 
\label{antit}
\end{eqnarray}
Expanding out the final state gives, for example,
\begin{eqnarray}
T^{11}_{22}&=&{\cal A}_3 \,{\cal T}^{11}_{22}\, {\cal A}_3=\frac{1}{6}
\Big\{{\cal T}^{11}_{22}\, {\cal A}_3 +\zeta\,\,{\cal T}^{21}_{12}\, {\cal A}_3
+\zeta\,\,{\cal T}^{31}_{22}\, {\cal A}_3
\nonumber\\
&& +\,\zeta\,\,{\cal T}^{11}_{32}\, {\cal A}_3+{\cal T}^{21}_{32}\, {\cal A}_3+     {\cal T}^{31}_{12}\, {\cal A}_3\Big\} \, , \label{anti}
\end{eqnarray}
corresponding to the six diagrams shown in Fig.~\ref{fig:symm}.  Each of these 6
diagrams generates 6 more terms when the initial state is symmetrized, for a total of
$6 \times 6=36$ terms in all.   Equation~(\ref{permsymm}) holds for both the initial
and final states, and because of this there is really only one distinct subamplitude
(all others are related to it by a phase), which we choose by convention to be 
${T}^{11}_{22}$.

\begin{figure}
\leftline{
\mbox{
\includegraphics[width=3.in]{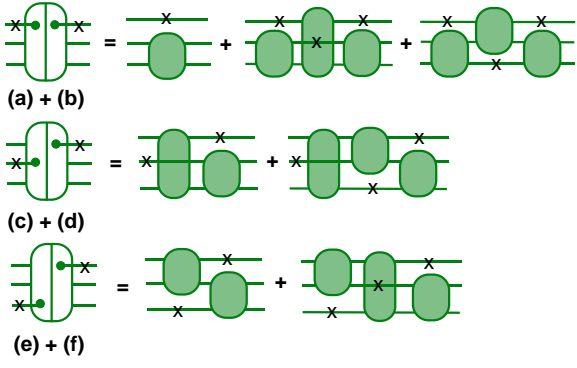}
}}
\caption{\footnotesize\baselineskip=10pt (Color online) 
Terms up to third order in the expansion of the
unsymmetrized subamplitudes shown in Fig.~\ref{fig:symm}. }
\label{fig:asymm}
\end{figure}  

\begin{figure}
\rightline{
\mbox{
\includegraphics[width=3.3in]{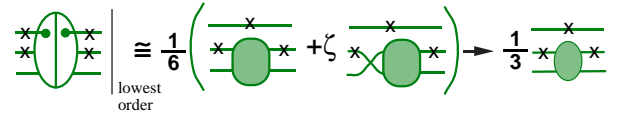}
}}
\caption{\footnotesize\baselineskip=10pt (Color online) 
The first order scattering can come from only two of the six
contributions shown in Fig.~\ref{fig:symm} [diagrams (a) and
(b)] and hence enters the symmetrized series with a weight of
$\frac{1}{3}$.  In this figure the oval (to the far right) is
the symmetrized two-body scattering amplitude, and the rounded
squares are the unsymmetrized two-body scattering amplitudes. }
\label{fig:first}
\end{figure}  

\begin{figure}
\leftline{
\mbox{
\includegraphics[width=3.3in]{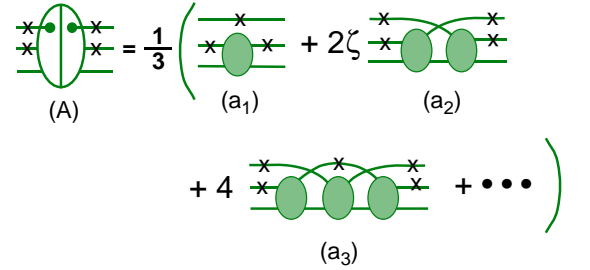}
}}
\caption{\footnotesize\baselineskip=10pt  (Color online) 
Final form of the series for the symmetrized subamplitude
$T^{11}_{22}$, with the proper weights for the contributions
from each order. }
\label{fig:seriesA}
\end{figure}  

\begin{figure}
\vspace*{-0.2in}
\rightline{
\mbox{
\includegraphics[width=3.3in]{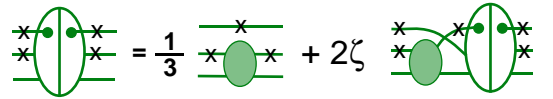}
}\hspace*{0.2in}}
\caption{\footnotesize\baselineskip=10pt  (Color online) 
Diagrammatic representation of the equation for the
spectator three-body scattering subamplitude ${\cal
T}^{11}_{22}$. This equation is equivalent to the series
shown in Fig. \ref{fig:seriesA}.}
\label{fig:scatteq}
\end{figure}

The symmetrization process introduces various weight factors
into the power series expansion of the amplitude
$T^{11}_{22}$, and hence into the equation for this amplitude. 
These were derived algebraically in Ref.~\cite{AGS97}; here
they are obtained diagrammatically from a study of
Figs.~\ref{fig:asymm}--\ref{fig:seriesA}.  Figure
\ref{fig:asymm} shows how each of the unsymmetrized
subamplitudes appearing in Fig.~\ref{fig:symm} is expanded up
to third order (in the two-body scattering). Considering the
symmetrization of the final state only, there are a total of
two first-order terms, four second-order terms, and 8 third-order 
terms.  The contributions from the first-order terms are
illustrated in Fig.~\ref{fig:first}; the end result is that the
symmetrized first-order scattering diagram enters with a
weight factor of $\frac{1}{3}$.  Applying the same argument to
the other terms gives the series shown in
Fig.~\ref{fig:seriesA}, which is unchanged by the
symmetrization of the initial state.  This series
results from the iteration of the scattering equation shown
diagrammatically in Fig.~\ref{fig:scatteq}, and hence this is
the correct scattering equation.  This is the result
obtained previously in Ref.~\cite{AGS97}.

\begin{figure}
%
\rightline{
\mbox{
\includegraphics[width=2.5in]{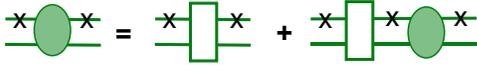}
}\hspace*{0.4in}}
\caption{\footnotesize\baselineskip=10pt  (Color online) 
Diagrammatic representation of the equation for the
symmetrized two-body scattering subamplitude $M^{1}_{22}$.
The open rectangle represents the irreducable kernel
$V^1_{22}$.}
\label{fig:Meq}
\end{figure}

In algebraic form, this scattering equation is      
\begin{eqnarray}
{T}^{11}_{22}=\frac{1}{3}M^1_{22}-2\zeta M^1_{22} G^1_2    
{\cal P}_{12} T^{11}_{22}\, , \label{scatteq}
\end{eqnarray} 
identical to Eq.~(3.55) of Ref.~\cite{AGS97}.  Here $M^1_{22}$
is the symmetrized amplitude for the two-body scattering of
particles 2 and 3 [with particle 1 a spectator], and $G^1_2$ is
the propagator for particle 3 off-shell.  The amplitude
$M^1_{22}$ satisfies the integral equation
\begin{eqnarray}
M^1_{22}&=&V^1_{22}-V^1_{22}\,G^1_2\,M^1_{22}\nonumber\\
&=&V^1_{22}-M^1_{22}\,G^1_2\,V^1_{22}\, ,
\label{Meq}
\end{eqnarray} 
where $V^1_{22}$ is the kernel, or driving terms, of the $NN$
interaction.  This equation is illustrated in
Fig.~\ref{fig:Meq}. In diagrams drawn in this paper, the minus
sign in the second term of (\ref{scatteq}) and (\ref{Meq}) will
be associated with the propagator $G^1_2$, so in the figures
the propagator is $-G^1_2$.  However, the factors
$\frac{1}{3}$ and
$2\zeta$ will be shown explicitly. 

The three-body bound state produces a pole in the $s=P^2$
channel (with $P^\mu$ the total momentum four-vector),
\begin{eqnarray}
{T}^{11}_{22}=-\frac{\left|\Gamma^1_2\right>
\left<\Gamma^1_2\right|}{M^2_t-P^2} + {\cal R}
\, ,
\label{poleeq}
\end{eqnarray} 
where the remainder term, ${\cal R}$, is regular at the pole,
and the spin structure of the propagating bound state is
included in the vertex function $\left|\Gamma^1_2\right>$. 
Substituting (\ref{poleeq}) into (\ref{scatteq}), approaching
the pole, and equating residues gives the three-body bound
state equation
\begin{eqnarray}
\left|\Gamma^1_2\right>=-2\zeta M^1_{22}
G^1_2     {\cal P}_{12} \left|\Gamma^1_2\right>\, .
\label{boundeq}
\end{eqnarray} 
Note that the inhomogenous term in Eq.~(\ref{scatteq}) has no
three-body bound state pole, and hence does not contribute to
the bound state equation.  The bound state equation is
therefore diagrammatically identical to Fig.~\ref{fig:scatteq}
without the inhomogenous term.

\begin{figure}
\vspace*{-0.2in}
\leftline{\hspace*{-0.1in}
\mbox{
\includegraphics[width=3.in]{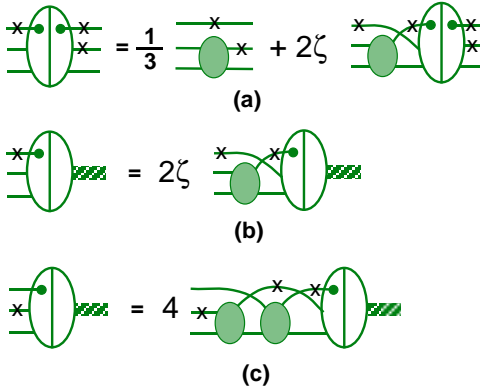}
}}
\caption{\footnotesize\baselineskip=10pt  (Color online) 
Diagrammatic representation of the definitions of 
three-body subamplitudes with {\it two\/} particles
off-shell. (a) The scattering amplitude with {\it both\/}
particles 2 and 3 off-shell, (b) the bound state vertex
function with particles 2 and 3 off shell, and (c) the bound
state vertex function with the spectator (particle 1) off
shell.}
\label{fig:offshell}
\end{figure}  

We conclude this section with a discussion of how to define
the spectator amplitudes when one of the spectators
is off shell.  The definitions needed in the subsequent
discussion are shown in Fig.~\ref{fig:offshell}.  Here the
principle is to expose the two-body interaction which
connects to the off-shell particle, because the two-body
amplitude can be extended off-shell by pulling out the last
two-body interaction (which is always defined with both
particles off shell).  Note that, when the spectator is
off-shell [Fig.~\ref{fig:offshell}(c)], the equation must be
iterated twice to get the desired result.

We now turn to the main subject of this paper, the
diagrammatic derivation of the three-body current operator.

\section{spectator three-body currents}

\subsection{The problem of double counting}

\begin{figure}
%
\leftline{
\mbox{
\includegraphics[width=3.1in]{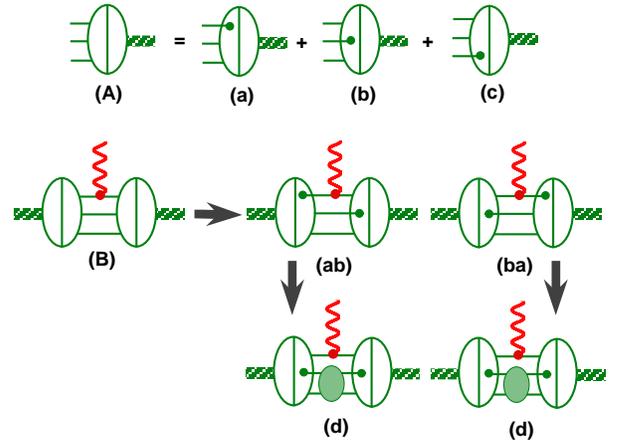}
}}
\caption{\footnotesize\baselineskip=10pt  (Color online) 
The upper panel shows the Bethe-Salpeter vertex function as a
sum of three subamplitudes for the three different choices of
the last interacting pair.  Lower panel shows that naive use
of this vertex function in the form factor gives overlap terms
(a)$\times$(b) and (b)$\times$(a) leading (after use of the
equation) to two terms of type (d), where only one should be
present. }
\label{fig:BSdouble}
\end{figure}  

To expose one of the central issues in the construction of
three-body currents, we begin by looking at what appears to be
the lowest order result in the Bethe-Salpeter formalism, and
show that this expected result leads to over counting.

The full BS three-body vertex function is the sum of three subamplitudes, as shown in
Fig.~\ref{fig:BSdouble}(A).  Guided by nonrelativistic theory, we might expect the
impulse approximation to the current to be related to the square of the wave
function, as illustrated in Fig.~\ref{fig:BSdouble}(B).  However, if this proposed
current is expanded using the wave equations, it leads to two terms of type
Fig.~\ref{fig:BSdouble}(d), while direct examination of the ladder sum (for example)
shows that there should be only one such term.  Unless an interaction term of type
(d) is explicitly subtracted from the ``impulse'' approximation, it will be double
counted.  The same problem does not arise in nonrelativistic theory because there the
diagrams represent a sequence of operators which, in general, do not commute. 
The iteration of (a)$\times$(b) gives a
different contribution from that of (b)$\times$(a), and both must be present.  [The
treatment of this problem in the context of the Bethe-Salpeter theory is discussed in
Ref.\ \cite{BK99}.]

It turns out that the spectator theory, like nonrelativistic
theory, also does not suffer from double counting.
Furthermore, the topology of the terms shown
in Fig.~\ref{fig:BSdouble} can be used to simplify the
spectator formalism.  A detailed demonstration of this is best
left until after the general results have been obtained below.


\subsection{Step I: Coupling photons to internal lines and
vertices}
\label{IIIB}

The current is constructed by coupling the photon to all
propagators and momentum dependent couplings in every diagram
in the infinite series.  This will give, according to a
general argument developed by Feynman, a conserved
current.  After the current has been constructed in this way, a
diagrammatic proof of current conservation will be given.

The construction of the current will be carried out in three
steps.  First the coupling to internal lines and vertices will
be constructed,  Then the wave equations will be used to
rearrange the result into a more usable form.  Finally,
the extension of the result to inelastic processes requires
coupling to the final state nucleons, and the correct way to
do this will be developed last.

\begin{figure}
%
\rightline{
\mbox{
\includegraphics[width=3.3in]{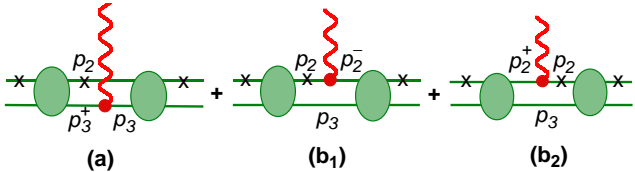}
}}
\caption{\footnotesize\baselineskip=10pt  (Color online) 
Diagrammatic representation of the gauging of the propagator
$G^1_2$ as given in Eq.~(\ref{gaugeprop}).  Diagrams (a),
(b$_1$), and (b$_2$) are the three terms on the r.h.s. of the
equation, respectively.    }
\label{fig:gauged-G}
\end{figure}  

The coupling to internal lines and vertices is very nicely
obtained using the gauging of equations method of Kvinikhidze
and Blankleider \cite{KB97}.  The first step in this
construction is to note that the coupling of the photon
satisfies the distributive rule of differential calculus. 
Starting from the scattering equation (\ref{scatteq}),
the photon coupling must satisfy the equation
%
\begin{eqnarray}
\left({T}^{11}_{22}\right)^\mu&=&\frac{1}{3}
\left(M^1_{22}\right)^\mu-2\zeta \left(M^1_{22}\right)^\mu
G^1_2{\cal P}_{12} T^{11}_{22}\nonumber\\&&
-2\zeta M^1_{22} \left(G^1_2\right)^\mu {\cal
P}_{12} T^{11}_{22}\nonumber\\&&
-2\zeta M^1_{22} G^1_2{\cal
P}_{12} \left({T}^{11}_{22}\right)^\mu ,\qquad
\label{gaugescatt}
\end{eqnarray} 
where $X^\mu$ denotes the coupling of the photon to all
internal lines and vertices in the series of Feynman diagrams
that make up $X$.  The gauging of the spectator propagator
generates three terms with the operators connecting each of
these terms having different arguments.  Denoting these
operators by ${\cal A}$ and ${\cal B}$, the gauged $G^1_2$ is
\begin{widetext}
\begin{eqnarray}
{\cal A}\left(G^1_2\right)^\mu{\cal B}&=&  
{\cal A}(p_2,p^+_3)\;
G(p_3^+)\,j_3^\mu(p_3^+,p_3)\;G(p_3)\,
(m+\not\! p_2)\;{\cal B}(p_2,p_3)\nonumber\\&&
+{\cal A}(p_2,p_3)\;(m+\not\! p_2)
\,j_2^\mu(p_2,p_2^-)\,G(p_2^-)\,G(p_3)
\;{\cal B}(p^-_2,p_3)\nonumber\\
&&+{\cal A}(p^+_2,p_3)\;G(p_2^+)\,j_2^\mu(p_2^+,p_2)
\,G(p_3)
(m+\not\! p_2)\;{\cal B}(p_2,p_3)\, .
\label{gaugeprop}
\end{eqnarray} 
\end{widetext}
In each term the particle with momentum $p_2$ is on shell, 
$p_2^2=m^2$, and $p_i^\pm=p_i\pm q$, and
\begin{eqnarray}
G(p)&=&(m-\not\!p)^{-1} \, .
\label{oneprop}
\end{eqnarray} 
Each term is a direct product of Dirac operators on the space
of particle 2 and 3 (with the space on which the operators act
implied by the momentum labels, so, for example, $\not\!p_2$
operates on the space of particle 2).  This equation is illustrated
diagrammatically in Fig.~\ref{fig:gauged-G}.  

\begin{figure}
%
\rightline{
\mbox{
\includegraphics[width=3.3in]{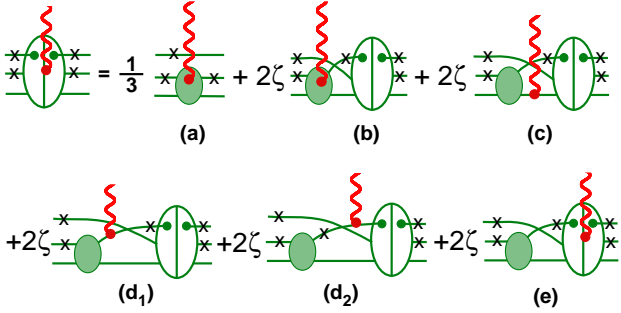}
}}
\caption{\footnotesize\baselineskip=10pt  (Color online) 
Diagrammatic representation of Eq.~(\ref{gaugescatt}), with (a)
the inhomogeneous term $\left(M^1_{22}\right)^\mu$, (b) the
term $\left(M^1_{22}\right)^\mu \!\!G^1_2{\cal P}_{12}
T^{11}_{22}$, (c), (d$_1$) and (d$_2$) the three terms
resulting from $(G^1_2)^\mu$ as illustrated in
Fig.~\ref{fig:gauged-G}, and (e) the term $M^1_{22}
G^1_2\,{\cal P}_{12} \left({T}^{11}_{22}\right)^\mu$.   }
\label{fig:internal-eq}
\end{figure}  

\begin{figure}
%
\rightline{
\mbox{
\includegraphics[width=3.2in]{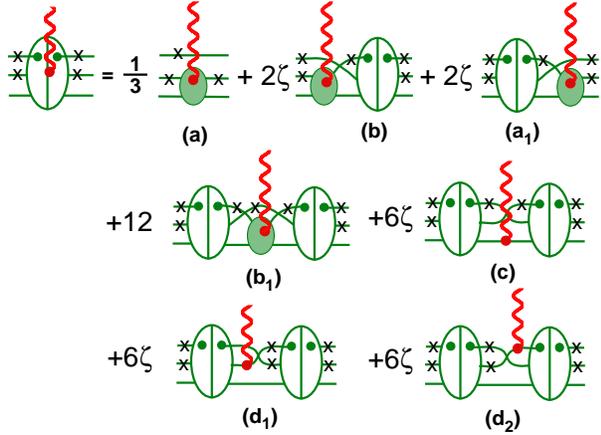}
}}
\caption{\footnotesize\baselineskip=10pt  (Color online) 
Diagrammatic representation of the solution to
Eq.~(\ref{gaugescatt}) for the current arrising from the
coupling of the photon to all {\it internal\/} lines and
vertices.  Diagrams (a$_1$) and (b$_1$) include effects of
rescattering to all orders arrising from diagrams (a) and
(b).  Similarly, diagrams (c), (d$_1$), and (d$_2$)
include higher order effects arrising from diagrams (c),
(d$_1$), and (d$_2$) of Fig.~\ref{fig:internal-eq}.  }
\label{fig:internal-sol}
\end{figure}  

\begin{figure}
%
\rightline{
\mbox{
\includegraphics[width=3.2in]{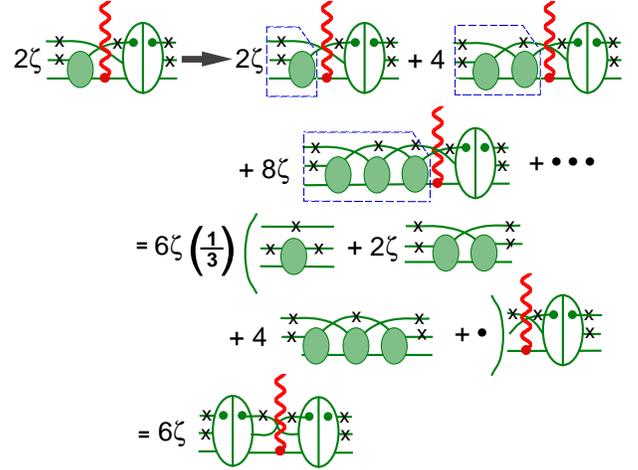}
}}
\caption{\footnotesize\baselineskip=10pt  (Color online) 
Figure showing how the the inhomogeneous term,
Fig.~\ref{fig:internal-eq}(c), is iterated by final state
interactions to all orders, leading to 
Fig.~\ref{fig:internal-sol}(c).  }
\label{fig:iteration}
\end{figure}  
 
Using Fig.~\ref{fig:gauged-G}, Eq.~(\ref{gaugescatt}) is
represented in Fig.~\ref{fig:internal-eq}.  This is an equation
for $\left({T}^{11}_{22}\right)^\mu$, and can be solved by
iteration, using the series representation for the three-body
scattering amplitude given in Fig.~\ref{fig:seriesA}.  The
solution is shown in Fig.~\ref{fig:internal-sol}.  To obtain
this solution diagrammatically is straightforward, if not
familiar to many.  Fig.\ \ref{fig:iteration} demonstrates
diagrammatically how Fig.~\ref{fig:internal-sol}(c) is
obtained by iterating the inhomogenous term in
Fig.~\ref{fig:internal-eq}(c).  
 
In subsequent applications the discussion will be limited to
those cases in which the initial state is bound.  These
diagrams are extracted from the general result shown in
Fig.~\ref{fig:internal-sol} by approaching the bound state
pole in the initial scattering and retaining the residue.
Diagrams (a) and (a$_1$) do not have such a pole, and
therefore do not contribute.  The result for the bound state
internal current is shown in Fig.~\ref{fig:internalbound-sol}.   

The parts of this figure involving rescattering in the final state are
identical to Fig.\ [4] of  KB \cite{KB97}.  To show this, first compare our
two-body scattering  Eq.\ (\ref{Meq}) with the KB two-body scattering equation
(Eq.\ (26) of Ref.\ \cite{KB97}). Note that $G^1_2=-\delta_2d_3$, so that the
equations are identical if $2 M^1_{22}= t_1$ and $2V^1_{22}= v_1$,
corresponding to a different normalization of the two-body amplitudes.  Next,
note that our three-body scattering Eq.\ (\ref{scatteq}) is identical to the
similar Eq.\ (15) of KB if we choose $\zeta=-1$ and set $6T^{11}_{22}=Xt_1$,
which corresponds to a different normalization of three-body scattering.  With
these replacements, our diagrams \ref{fig:internalbound-sol} ($b_1$), ($c$),
and ($d$) are identical to Fig.\ [4] in KB [recalling that second figure in the
KB Fig.\ [4] is the same as the sum of our (d$_1$) and (d$_2$)].   Since the KB
result shown in Fig.\ [4] applies only to transitions to connected final
states, diagrams like our Fig.\ \ref{fig:internalbound-sol} (b) will be
discussed after we have finished our discussion of the breakup process.

Note that Fig.~\ref{fig:internalbound-sol}(d$_2$) includes a contribution in
which the spectator (particle 1) is off-shell.   While recognizing that further
reductions are possible, KB elect to leave their answer in this form.    In
Step II of our derivation, we use the wave equations (\ref{scatteq}) and
(\ref{boundeq}) to replace this amplitude by an equivalent one in which the
spectator is on-shell.   This replacement does not change the total result, but
is still very useful for numerical applications, and leads to a nice
demonstration of how the spectator equations avoid the double counting problem
in a natural way.

\subsection{Step II: Removal of off-shell spectator
contributions}

\begin{figure}
%
\leftline{
\mbox{
\includegraphics[width=3.1in]{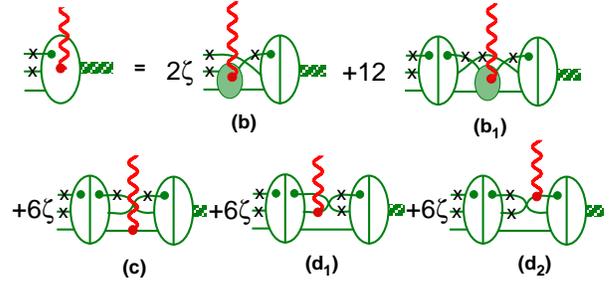}
}}
\caption{\footnotesize\baselineskip=10pt  (Color online) 
Diagrammatic representation of the solution,
Fig.~\ref{fig:internal-sol}, applied to processes with an
incoming three-body bound state.  }
\label{fig:internalbound-sol}
\end{figure}  

\begin{figure}
%
\leftline{\hspace{-0.3in}
\mbox{
\includegraphics[width=3.1in]{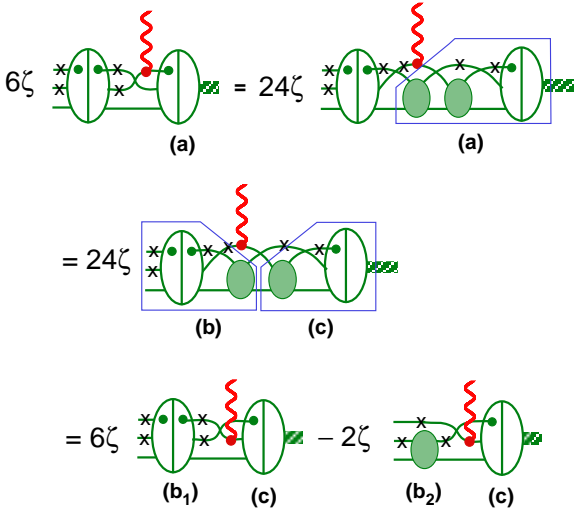}
}}
\caption{\footnotesize\baselineskip=10pt  (Color online) 
Diagrammatic representation of the transformation of diagram
(d$_2$) in Fig.~\ref{fig:internalbound-sol}.  Ths first step
uses Fig.~\ref{fig:offshell}(c) to replace the initial three
body amplitude [denoted by (a) in the figure].  Then the two
parts labeled (b) and (c) are isolated, and
Fig.~\ref{fig:offshell}(a) is used to replace (b) and 
Fig.~\ref{fig:offshell}(b) to replace (c).  The final result
is the two terms shown in the lower panel.  Note that the diagram (b$_2$)(c) cancels an identical diagram that comes from the first iteration in the final state scattering amplitude in (b$_1$)(c) [c.f.  diagram (a$_1$) of Fig.\ \ref{fig:seriesA}].}
\label{fig:replace}
\end{figure}  

The diagram (d$_2$) of Fig.~\ref{fig:internalbound-sol} can
be transformed as shown in Fig.~\ref{fig:replace}.  The first
step is to replace the initial bound state amplitude (which
has its spectator off-shell) using the definition shown in
Fig.~\ref{fig:offshell}(c).  Then we recognize that, {\it if\/}
one of the two $NN$ scattering amplitudes introduced by this
substitution is identified with the final state, and one
with the initial state (as outlined by the boxes in the
figure), it is possible to use the equations illustrated in
Fig.~\ref{fig:offshell}(a) and (b) to further simplify the
diagram.  The result is the two diagrams shown in the bottom
panel.  The conclusion is that diagram (d$_2$) can be
replaced by these two diagrams, {\it neither of which has
the spectator off-shell\/}.

\begin{figure}
%
\leftline{\hspace{-0.1in}
\mbox{
\includegraphics[width=3.2in]{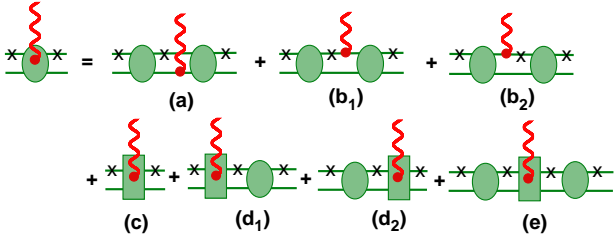}
}}
\caption{\footnotesize\baselineskip=10pt  (Color online) 
Diagrammatic representation of the gauged $(M^1_{22})^\mu$
(where, in our application, the on-shell particle is \#2 and
the off-shell particle \#3).  Figures (b$_1$) and
(b$_1$) arise from the gauging of the propagator
$(G^i_j)^\mu$, and (c)-(e) include contributions from the
two-body interaction current, represented by
the photon coupling inside a shaded rectangle.  There are no
three-body forces, and hence no three-body interaction
currents.}
\label{fig:NNgauge}
\end{figure}  

\begin{figure}
%
\leftline{
\mbox{
\includegraphics[width=3.0in]{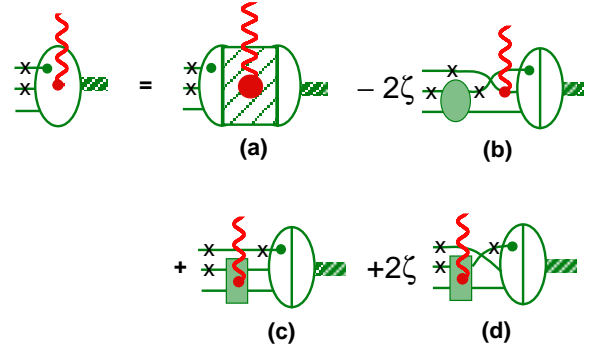}
}}
\caption{\footnotesize\baselineskip=10pt  (Color online) 
Diagrammatic representation of contributions from internal
photon couplings to the three-body breakup process.  Diagram
(a) is constructed from the core diagrams shown in
Fig.~\ref{fig:core} (simply add the two external half ellipses
to each diagram in Fig.~\ref{fig:core} to complete the
diagram).  Diagram (b) is the same diagram (b$_2$)(c) that appeared in 
Fig.\ \ref{fig:replace} when the off-shell spectator was replaced by 
an on-shell spectator.  It cancels the first iteration of the final state 
interaction in Fig.~\ref{fig:core}(e$_2$) so that the sum of
all of these contributions contains {\it no\/} couplings to external
nucleons. }
\label{fig:internalfinal}
\end{figure}  

\begin{figure}
\rightline{
\mbox{
\includegraphics[width=3.5in]{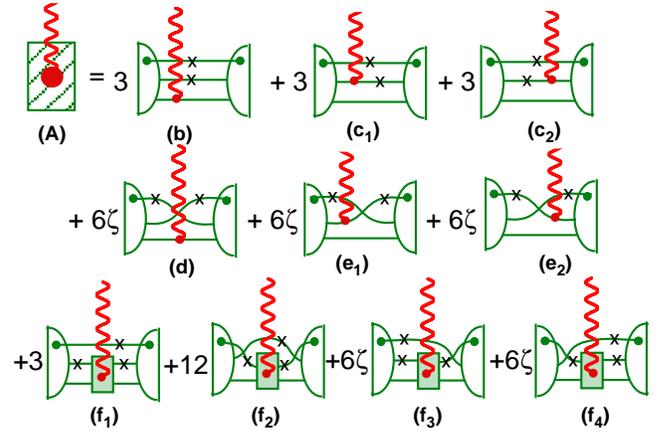}
}\hspace*{-0.1in}}
\caption{\footnotesize\baselineskip=10pt  (Color online) 
This figure shows ``core'' diagrams common to all
interactions.  The half circles in the initial and final
state can be either the bound state, or part of a three-body
scattering amplitude.  Substituting into
Fig.~\ref{fig:internalfinal} gives the complete
internal current. (A) the core equals the sum of (b) and (d)
photon coupling to off shell particle, either with or without
exchange of the 2 on-shell particles, (c) and (e) the coupling
of the photon to the on-shell interacting particle (arranged
so that the spectator is always on-shell), and (f)
two-body interaction current diagrams. }
\label{fig:core}
\end{figure} 
  
\begin{figure}
\begin{center}
\mbox{
\includegraphics[width=2in]{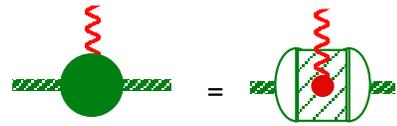}
}
\end{center}
\caption{\footnotesize\baselineskip=10pt  (Color online) 
The three-body bound state form factor is constructed only
from the core diagrams of Fig.~\ref{fig:core}. }
\label{fig:ff}
\end{figure} 

It is convenient to further simplify
Fig.~\ref{fig:internalbound-sol} by replacing the interaction
of the photon internal to the $NN$ amplitude with the gauged
result for the $NN$ amplitude,
\begin{eqnarray}
(M^1_{22})^\mu= (V^1_{22})_{\rm I}^\mu +(V^1_{22})_{\rm I}^\mu\,G^1_2\,
M^1_{22} + M^1_{22}\,G^1_2\,(V^1_{22})_{\rm I}^\mu\nonumber\\
+M^1_{22}\,G^1_2\,(V^1_{22})_{\rm I}^\mu\,G^1_2\,M^1_{22}
+M^1_{22}\,(G^1_{2})^\mu\,
M^1_{22}\, ,\quad
\label{NNgauge}
\end{eqnarray} 
illustrated in Fig.~\ref{fig:NNgauge}.  This
replacement exposes the two-body interaction current,
$(V^1_{22})_{\rm I}^\mu$ which includes the photon coupling to all
exchanged mesons and meson-nucleon vertices (when present). 
Using this expansion, and the wave equations, gives the
diagrams shown in Fig.~\ref{fig:internalfinal}, with the
``core'' diagrams defined in Fig.~\ref{fig:core}.  The core
diagrams also define the three-body form factor.  To extract
this form factor from the result shown in
Fig.~\ref{fig:internalfinal}, go to the bound-state pole in the
final scattering amplitude, and note that the diagrams (b),
(c) and (d) will not contribute.  The result is summarized in
Fig.~\ref{fig:ff}.

It remains now to (i) demonstrate that the form factor
conserves current, (ii) find the additional diagrams
(describing couplings to external nucleons) that must be added
to Fig.~\ref{fig:internalfinal} to compete the
description of the three-body breakup process, and
(iii) show that the three-body breakup also conserves
current.  

It turns out that the best way to proceed with these remaining
tasks is to first derive the WT identity for the core
diagrams, Fig.~\ref{fig:core}.

\subsection{Step III: The WT identity for the core
diagrams}
 
Our study of current conservation is based on a generalization
of the arguments of Gross and Riska \cite{GR87}.  There it was
shown that the current will be conserved if (i) it is
constructed from elementary nucleon and interaction currents
that satisfy the appropriate WT identities, and (ii)
contributions from {\it all possible\/} couplings of the photon
to nucleons and interactions are included in a consistent
manner.  The core diagrams derived in the previous subsection
provide a consistent scheme for coupling photons to all two-body 
interactions and {\it internal\/} nucleons, and
hence provide the solution to condition (ii) when there are no
external free nucleons (true for the form factor, but not for
the three-body breakup).  It remains now to show explicitly how
the WT identities for the elementary nucleon and interaction
currents insure that this is so.  

\begin{figure}
\begin{center}
\mbox{
\includegraphics[width=3.3in]{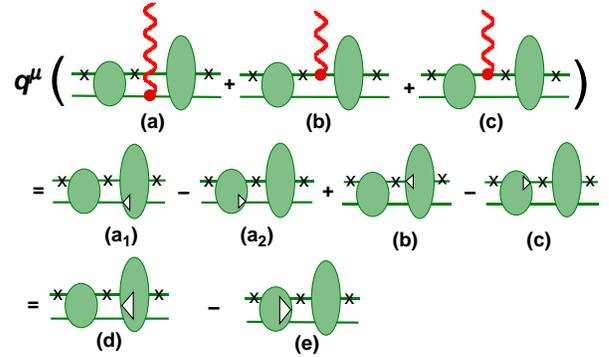}
}
\end{center}
\caption{\footnotesize\baselineskip=10pt  (Color online) 
Diagrammatic representation of Eq.~(\ref{WTonGij}).  The
small open triangle on each amplitude marks where the photon
was inserted, so the momentum of the amplitude at the location
of the triangle does not match the momentum of the propagator
connected to it.  The larger triangle denotes the sum of two
contributions: (d) is the sum of (a$_1$) and (b); (e) the sum
of (a$_2$) and (c). }
\label{fig:WTonG}
\end{figure} 

\begin{figure}
\leftline{\hspace{-0.3in}
\mbox{
\includegraphics[width=3.6in]{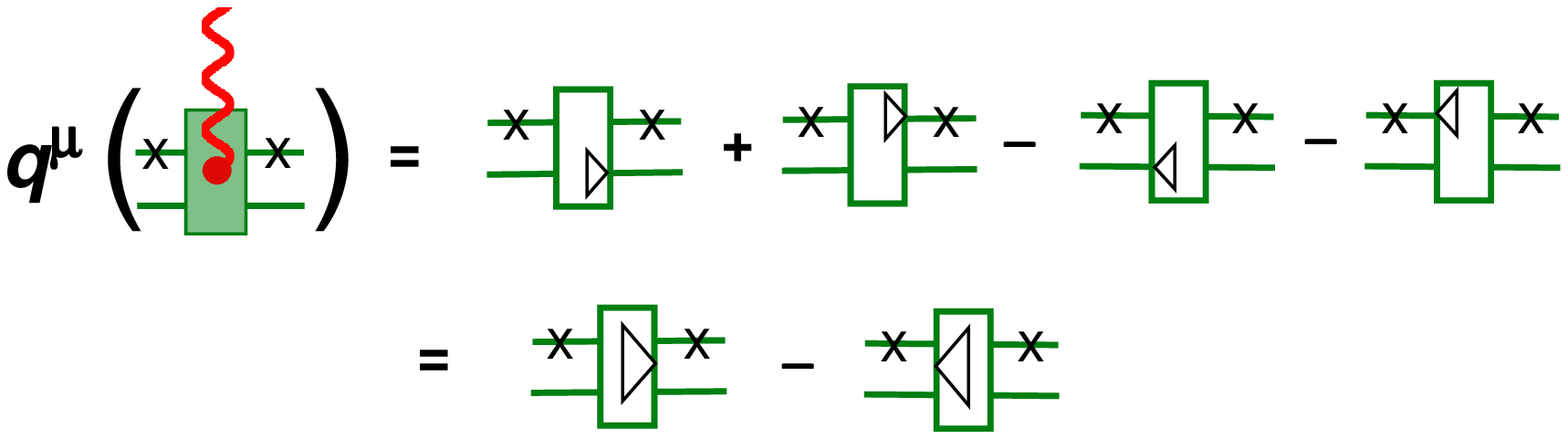}
}
}
\caption{\footnotesize\baselineskip=10pt  (Color online) 
Diagrammatic representation of Eq.~(\ref{WTonV}).  The
meaning of the open triangles is explained in
Fig.~\ref{fig:WTonG}. }
\label{fig:WTonV}
\end{figure}

The nucleon current is
constructed to satisfy the WT identity
\begin{eqnarray}
q_\mu j_i^\mu(p',p)=\left[G^{-1}(p)-G^{-1}(p')\right] \, ,
\label{WTiden}
\end{eqnarray} 
where the particle charge is excluded from the
definition of the current, and  conservation of four-momentum
at every vertex implies that
$q=p'-p$.  From this it follows that the divergence of
Eq.~(\ref{gaugeprop}) is  
\begin{eqnarray}
&&q_\mu{\cal A}\left(G^1_2\right)^\mu{\cal B} 
=\nonumber\\
&&\qquad {\cal A}(p_2,p_3^+)\;
(m+\not\!p_2)\,G(p^+_3)\;{\cal B}(p_2,p_3)
\nonumber\\
&&\qquad-{\cal A}(p_2,p_3^+)\;
(m+\not\!p_2)\,G(p_3) \;{\cal B}(p_2,p_3)
\nonumber\\
&&\qquad+{\cal A}(p_2,p_3)\;(m+\not\!p_2)
G(p_3)\;{\cal B}(p^-_2,p_3)\nonumber\\
&&\qquad-{\cal A}(p^+_2,p_3)\;G(p_3) (m+\not\!p_2)
\;{\cal B}(p_2,p_3) \, .
\qquad
\label{WTonGij}
\end{eqnarray}
This relation is illustrated diagrammatically in
Fig.~\ref{fig:WTonG}.  Note that Fig.~\ref{fig:WTonG}(d) is a
shorthand notation for the 1st and 3rd terms on the r.h.s.\
of Eq.~(\ref{WTonGij}) and Fig.~\ref{fig:WTonG}(e) for the
2nd and 4th terms.  
Similarily, the WT identity satisfied by the interaction
current (for further discussion, see Eq.~(3.3) of \cite{GR87})
is  
\begin{eqnarray}
q_\mu\left(V^1_{22}\right)_{\rm I}^\mu\!\!\! =\! \left[
V^1_{22}(p'_2,p'^-_3;p_2,p_3)-
V^1_{22}(p'_2,p'_3;p_2,p^+_3)\right]\nonumber\\
+\left[V^1_{22}(p'^-_2,p'_3;p_2,p_3)-
V^1_{22}(p'_2,p'_3;p^+_2,p_3)\right]
\, ,\qquad
\label{WTonV}
\end{eqnarray}
with $p_i^\pm$ defined above.  In this
equation, the photon momentum is inserted wherever there is a
$p^\pm_i$, as in Eq.~(\ref{WTonGij}).  Using the notation of
Fig.~\ref{fig:WTonG}, this equation is illustrated in
Fig.~\ref{fig:WTonV}.

\begin{figure}
\vspace*{-0.3in}
\rightline{
\mbox{
\includegraphics[width=3.3in]{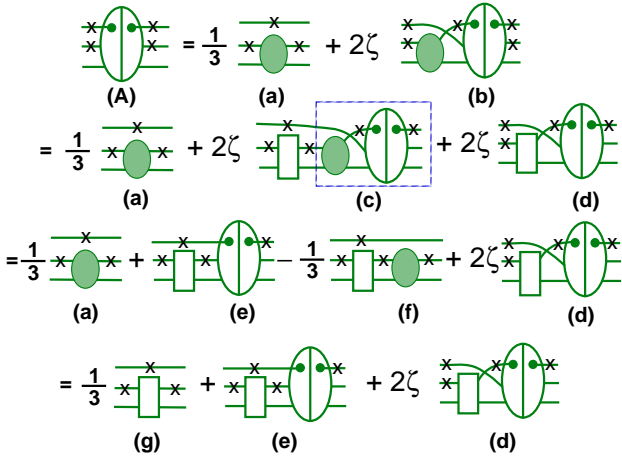}
}}
\caption{\footnotesize\baselineskip=10pt  (Color online) 
Diagrammatic derivation of the replacement of the scattering
amplitude by expressions involving the interaction kernel
$V^1_{22}$.  The top line shows the wave equation with the
amplitude (A) equal to the inhomogeneous part (a) and the
rescattering part (b).  In the second line, $M^{11}_{22}$ in
(b) is replaced by its scattering equation, Fig.~\ref{fig:Meq},
and in the third line one of these terms, (c), is replaced by
the three-body scattering equation. Because of the
cancellation of diagrams (a) and (f), the final result (last
line) is the sum of only three diagrams, (g), (e) and (d).  The
diagrams (a), (f), and (g) do not contribute to bound states,
leaving only diagrams (e) and (d).}
\label{fig:Vpullout}
\end{figure}  

\begin{figure}
%
\leftline{\hspace{-0.3in}
\mbox{
\includegraphics[width=3.3in]{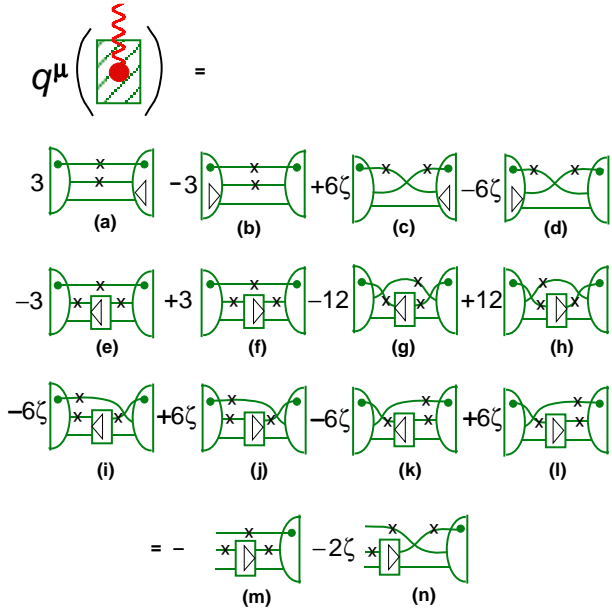}
}}
\caption{\footnotesize\baselineskip=10pt  (Color online) 
Derivation of the WT identity for the core diagrams {\it when 
connected to an initial bound state and a final scattering
state\/}. Using Fig.~\ref{fig:WTonG}, the divergence of core
diagrams
\ref{fig:core}(b)--\ref{fig:core}(e) gives diagrams (a) through
(d) above, and using Fig.~\ref{fig:WTonV} the divergence of
the interaction current diagrams \ref{fig:core}(f) gives
diagrams (e) through ($l$).  These diagrams reduce to (m) and
(n), as discussed in the text.} 
\label{fig:WToncore}
\end{figure}  

The derivation of the WT identity for the core diagrams depends
on the observation that the three-body bound state and
scattering equations can be used to express the amplitudes in
terms of the two-body kernel
$V^1_{22}$, as illustrated in Fig.~\ref{fig:Vpullout}.  Using
this relation, and the WT identities (\ref{WTonGij}) and
(\ref{WTonV}), the WT identity for the core diagrams can be
found.  The steps are outlined in
Fig.~\ref{fig:WToncore}.  Cancellations occur
when the identitities of Fig.~\ref{fig:Vpullout} are used to
reexpress the diagrams
\ref{fig:WToncore}(b)--\ref{fig:WToncore}(e).  In detail,
\ref{fig:WToncore}(a) cancels \ref{fig:WToncore}(e) and
\ref{fig:WToncore}(i), \ref{fig:WToncore}(b) cancels
\ref{fig:WToncore}(f) and \ref{fig:WToncore}(j) leaving
\ref{fig:WToncore}(m),
\ref{fig:WToncore}(c) cancels
\ref{fig:WToncore}(g) and \ref{fig:WToncore}(k), and 
\ref{fig:WToncore}(d) cancels
\ref{fig:WToncore}(h) and \ref{fig:WToncore}($l$) leaving
\ref{fig:WToncore}(n). 

This result implies that the bound-state form factor
conserves current, because in this case neither of the
diagrams \ref{fig:WToncore}(m) or \ref{fig:WToncore}(n) are
present, and the WT identity gives zero.  Discussion of the
three-body breakup process requires additional diagrams, which
will be discussed now. 

\subsection{Step IV: Photon coupling to external nucleons}

\begin{figure}
%
\rightline{
\mbox{
\includegraphics[width=3.4in]{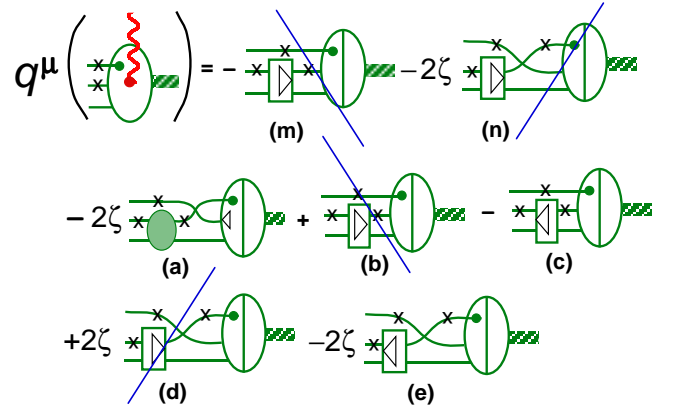}
}}
\caption{\footnotesize\baselineskip=10pt  (Color online) 
Derivation of the WT identity for the internal photon 
couplings to the three-body breakup process
[Fig.~\ref{fig:internalfinal}]. Using Fig.~\ref{fig:WToncore},
the divergence of core diagrams leaves diagrams (m) and (n). 
The divergence of
\ref{fig:internalfinal}(b) gives diagram (a), the divergence
of \ref{fig:internalfinal}(c) gives (b) and (c), and the
divergence of diagram \ref{fig:internalfinal}(d) gives (d) and
(e).  Diagram (m) is cancelled by (b) and diagram (n) is
cancelled by (d), leaving the three diagrams (a), (c), and
(e).} 
\label{fig:WToninternal}
\end{figure}

Using the results of Fig.~\ref{fig:WToncore}, we obtain the WT
identity for the internal photon couplings to the three-body
breakup process, shown in Fig.~\ref{fig:internalfinal}.  The
result is shown in Fig.~\ref{fig:WToninternal}. 

\begin{figure}
%
\leftline{
\mbox{
\includegraphics[width=3.3in]{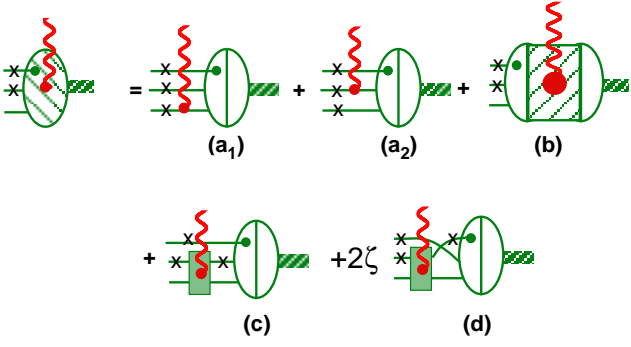}
}}
\caption{\footnotesize\baselineskip=10pt  (Color online) 
Diagrams that describe the three-body breakup process.
Diagrams (a) are the RIA, (b) the FSI, and (c) and (d) are
interaction currents (I).} 
\label{fig:breakup}
\end{figure}

The coupling to external nucleons will produce the terms needed to cancel the
diagrams in Fig.~\ref{fig:WToninternal}(a), \ref{fig:WToninternal}(c), and
\ref{fig:WToninternal}(e).  Since \ref{fig:WToninternal}(a) results only from
Fig.~\ref{fig:internalfinal}(b), \ref{fig:internalfinal}(b) will be removed
from the final three-body breakup current.  [Removal of this diagram means that
the core contributions will contain a diagram like
Fig.~\ref{fig:internalfinal}(b) that can be interpreted as an interaction with
a free final state particle, and care must be taken not to overlook this term. 
We will discuss this further in the conclusions.]  The final result for
three-body breakup, including photon coupling to external nucleons, is shown in
Fig.~\ref{fig:breakup}.

\begin{figure}
%
\rightline{
\mbox{
\includegraphics[width=3.4in]{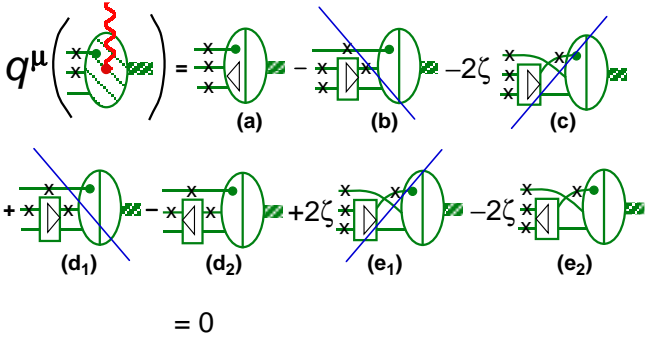}
}}
\caption{\footnotesize\baselineskip=10pt  (Color online) 
Proof that the breakup diagrams conserve current.  Diagram
(a) results from application of the WT identity
(\ref{WTonGij}) on the RIA terms (with the final
state on shell), (b) and (c) are from
Fig.~\ref{fig:WToncore}, and (d) and (e) result from the the
application of Fig.~\ref{fig:WTonV} on the I diagrams
\ref{fig:breakup}(c) and \ref{fig:breakup}(d), respectively. 
After the cancellations shown, the application of
Fig.~\ref{fig:Vpullout} to diagram (a) cancels (d$_2$) and
(e$_2$), giving zero. } 
\label{fig:WTonbreakup}
\end{figure}

The three-body breakup diagrams (a$_1$) and (a$_2$) will be
referred to as the relativistic impulse approximation (RIA),
diagrams (c) and (d) as interaction currents (denoted by I,
but not to be confused with isobar currents which, if
present, are included in the interaction currents), and the
core contribution (b), which includes final state
interactions driven by both the RIA and the I (and denoted FSI).  
A proof that this set of diagrams
conserves current is given in Fig.~\ref{fig:WTonbreakup}.  

We have completed our derivation of the current, and return
now to the discussion of double counting first introduced in
the last section.

\subsection{Conclusion: Removal of double counting in the
Spectator theory}

We now demonstrate that the problem of double counting
referred to in subsection A above is solved by the current
operator given in Fig.~\ref{fig:breakup}. 

\begin{figure}
%
\rightline{
\mbox{
\includegraphics[width=3.4in]{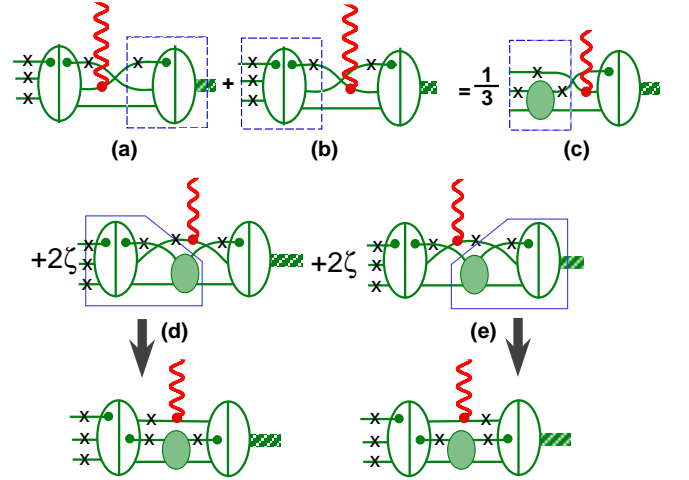}
}}
\caption{\footnotesize\baselineskip=10pt  (Color online) 
Diagrams (e$_1$) and (e$_2$) of Fig.~\ref{fig:core} [redrawn
here as (a) and (b) and without their overall factor of 6$\zeta$] are equal to (c) + (d) + (e). The last line redraws (d) and (e) so that they can be easily compared
with the diagrams shown in Fig.~\ref{fig:BSdouble}. } 
\label{fig:SpecDouble}
\end{figure}

Guided by Fig.~\ref{fig:BSdouble}, we examine the ``exchange''
terms \ref{fig:core}(e) that contribute
to the core process.  These diagrams are reproduced in
Figs.~\ref{fig:SpecDouble}(a) and (b) for the case when the
initial state is bound and the final state is three-body
breakup.  Using the bound state wave equation with
Fig.~\ref{fig:SpecDouble}(a)  and the scattering equation
with Fig.~\ref{fig:SpecDouble}(b) gives the three
contributions (c), (d), and (e).  Figures (d) and (e) are
untangled in the last line of the figure, so that they may be
more easily compared with diagrams \ref{fig:BSdouble}(d) of
the Bethe Salpeter theory.  Note that in the spectator theory,
{\it both diagrams (d) and (e) must occur\/}, because they
describe different processes, with the spectator on shell
either ``before'' or ``after'' the interaction (note that here
``before'' or ``after'' refer to a topological ordering and
not a time ordering).  Our insistence that the spectator always
be on shell has eliminated the double counting problem.

\begin{figure}
%
\leftline{\hspace{-0.1in}
\mbox{
\includegraphics[width=3.1in]{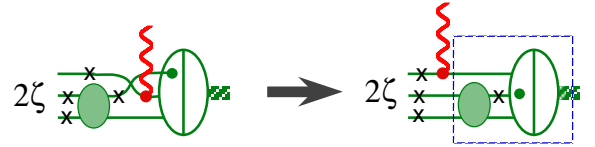}
}}
\caption{\footnotesize\baselineskip=10pt  (Color online) 
Fig.~\ref{fig:SpecDouble}(c) (multiplied by the missing factor of 6$\zeta$) is redrawn showing how it could
be interpreted in a different way. } 
\label{fig:interpretfinal}
\end{figure}

Finally, look at diagram \ref{fig:SpecDouble}(c), which arises
from final state interactions.  In
Fig.~\ref{fig:interpretfinal} this diagram is rearranged to
look like an RIA contribution with the spectator {\it off
shell\/}.  This would lead to double counting if we had
included such processes in the RIA, but these contributions
are explicitly excluded from the RIA contributions (a$_1$) and
(a$_2$) shown in Fig.~\ref{fig:breakup}.

We now record, for future use, the algebraic espressions
corresponding to our major results.

\begin{figure}
\vspace*{0.2in}
\centerline{
\mbox{
\includegraphics[width=2.5in]{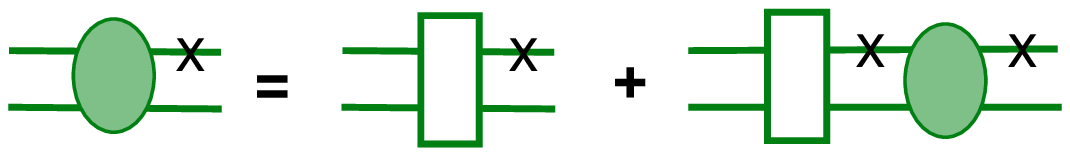}
}}
\caption{\footnotesize\baselineskip=10pt  (Color online) 
Diagrammatic representation of the equation for the
symmetrized two-body scattering subamplitude $M^{1}_{02}$ with
{\it both\/} final particles off shell.}
\label{fig:Meqoff}
\end{figure}  

\begin{figure}
%
\rightline{
\mbox{
\includegraphics[width=3.7in]{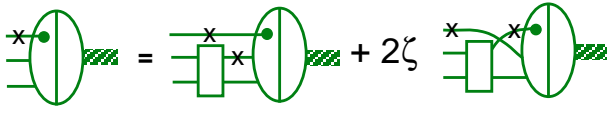}
}}
\caption{\footnotesize\baselineskip=10pt  (Color online) 
Diagrammatic representation of the equation for the
three-body vertex function with {\it both\/} interacting
particles off-shell.}
\label{fig:offshell2}
\end{figure}  

\section{Algebraic expressions for the wave functions and
currents}

In this section we record the algebraic form of the three-body vertex
functions, wave functions, form factors, and three-body breakup.  The Appendix
shows in detail how the covariant three-body normalization condition leads to
the conservation of charge.

\subsection{The three-body vertex function and wave function}

The three-body vertex function defined in Ref.~\cite{SGF97}
[Eq.~(3.14)] is denoted  
\begin{eqnarray}
\Gamma_{\lambda_1\lambda_2\alpha}(k_1,k_2,k_3)\equiv
\left<k_1\lambda_1(k_2\lambda_2\,
k_3\alpha)|\Gamma^1_2\right>\, ,  \label{vertexrest}
\end{eqnarray}
where $\left<k_1\lambda_1(k_2\lambda_2\,
k_3\alpha)|\Gamma^1_2\right>$ is the three-body vertex
function  describing the coupling of the $^3$He nucleus to
the three-nucleon system, with the first pair of arguments
[$k_1, \lambda_1$] the  four-momentum and helicity of the
on-shell spectator, the second pair
[$k_2, \lambda_2$] the  four-momentum and helicity of the
on-shell particle in the interacting pair, and the last pair
[$k_3, \alpha$] the four-momenta and Dirac index of the
off-shell particle in the interacting pair.  Different
notations will sometimes be used for these momenta; it
is their location in the argument list that identifies them as
spector, on-shell interacting particle, or off-shell
particle.  In this notation the momenta and spin indicies of
all three particles are defined in the same system, with 
\begin{eqnarray}
P=k_1+k_2+k_3  \label{Ptotal}\, ,
\end{eqnarray}
the total four momentum of the bound state.  In the {\it rest
frame\/} of three-body bound state, the energy of the
off-shell particle is 
\begin{eqnarray}
k^0_3=M_B-E_1-E_2 <m  \label{offshellE}\, .
\end{eqnarray}
In this representation, the symmetry of the amplitude is
simply   
\begin{eqnarray}
\Gamma_{\lambda_1\lambda_2\alpha}(k_1,k_2,k_3)&=&\zeta\;
{\cal P}_{12}\Gamma_{\lambda_1\lambda_2\alpha}(k_1,k_2,k_3)
\nonumber\\
&=&
\zeta\;\Gamma_{\lambda_2\lambda_1\alpha}(k_2,k_1,k_3) \, .
\label{vertexsymm}
\end{eqnarray}

When solving the three-body bound state equations (\ref{boundeq}) it is
convenient to work in the rest frame of the bound state, but to boost the
interacting pair to its own rest frame, where the partial wave decomposition of
the two-body  amplitude that drives the equation,  $M^1_{22}$, is defined. 
The  numerical solutions reported in Ref.~\cite{SG97} were carried out in this
mixed frame,  {\it i.e., with $k_1$ and $\lambda_1$ defined in the three-body
rest system and the remaining variables defined in the rest system of the
interacting 23 pair\/}.  The connection between these two representations will
be discussed in a subsequent paper, where we will calculate the
electrodisintegration of $^3$He \cite{GPS}.   In this section we will use the
representation (\ref{vertexrest}).  

The wave function, when needed, is defined by
\begin{eqnarray}
\Psi_{\lambda_1\lambda_2\alpha}(k_1,k_2,k_3)=
G_{\alpha,\alpha'}(k_3)
\Gamma_{\lambda_1\lambda_2\alpha'}(k_1,k_2,k_3)
\, ,
\label{wavefunction}
\end{eqnarray}
where the nucleon propagator, $G$, was given in
Eq.~(\ref{oneprop}).


\subsection{The three-body form factors} 

The diagrams needed to calculate the form factors were
displayed in Figs.~\ref{fig:core} and \ref{fig:ff}.  Following
work on the deuteron form factors \cite{VODG95}, the
diagrams Figs.~\ref{fig:core}(b)--(e) are referred to as 
the complete impulse approximation (CIA).  Diagrams
Figs.~\ref{fig:core}(f) are the interaction currents, denoted
by I.

Some of the CIA diagrams require knowledge of the vertex
function with the {\it two interacting\/} nucleons
off-shell.   This vertex function was defined in
Fig.~\ref{fig:offshell}(b).  A more convenient expression for
this vertex function can be found using the equation for
the two-body scattering amplitude with {\it both\/} particles
in the final state off-shell [this generalization of
Fig.~\ref{fig:Meq} is shown in Fig.~\ref{fig:Meqoff}]. 
Substituting Fig.~\ref{fig:Meqoff} into
Fig.~\ref{fig:offshell}(b), and using
Fig.~\ref{fig:offshell}(b) a second time, gives the result
shown in Fig.~\ref{fig:offshell2} for the bound-state
amplitude with particles 2 and 3 off-shell.  [This is a
generalization of a result previously shown in
Fig.~\ref{fig:Vpullout}.]  The algebraic form of the equation
shown in Fig.~\ref{fig:offshell2} is
\begin{widetext}
\begin{eqnarray}
\Gamma_{\lambda_1\beta\alpha}(k_1,k_2,k_3)=-\int
\frac{m\,d^3k'_2}{E'_2\,(2\pi)^3}
\sum_{\lambda_2}
V_{\beta\alpha,\lambda_2\alpha'}(k_2,k_3;k_2',k_3')\,
[1+2\,\zeta\,{\cal P}_{12}]\,
\Psi_{\lambda_1\lambda_2\alpha'}(k_1,k'_2,k'_3)
\, ,
\label{vertexoffshell}
\end{eqnarray}
where $V$ is the same two-body kernel used in
Eq.~(\ref{Meq}), and summation over repeated Dirac
indicies is implied.  The phase of each spectator amplitude is
computed from $i(-i)^n$, where $n$ is the total number of
off-shell propagators plus interactions (either vertices or
kernels) in the integrand of the amplitude.   Equation
(\ref{vertexoffshell}) has one off-shell propagator
(contained in $\Psi$), and two interactions ($V$ and
$\Gamma$), for a phase of $i(-i)^3=-1$.   For convenience we
have adopted the notation
\begin{eqnarray}
V_{\beta\alpha,\beta'\alpha'}(k_2,k_3;k_2',k_3')\,
u_{\beta'}(k_2',\lambda'_2)&\equiv&
V_{\beta\alpha,\lambda'_2\alpha'}(k_2,k_3;k_2',k_3')
\nonumber\\
\bar u_{\beta}(k_2,\lambda_2)
V_{\beta\alpha,\lambda_2\alpha'}(k_2,k_3;k_2',k_3')&\equiv&
V_{\lambda_2\alpha,\lambda'_2\alpha'}(k_2,k_3;k_2',k_3')\, ,
\label{vnotation}
\end{eqnarray}
so care must be taken to distinguish Dirac indicies from helicity indices. 
Whenever a Dirac index is replaced by a helicity index, a contraction with an
on-shell, positive energy spinor, such is shown in Eq.~(\ref{vnotation}), is
implied.  The on-shell Dirac spinors are normalized to $\bar u u=1$.   

The algebraic result for
the six diagrams that make up the CIA can now be written
\begin{eqnarray}
J^\mu_{\rm CIA}&=&3e\int\!\!\int
\frac{m^2\,d^3k_1d^3k_2}{E_1E_2\,(2\pi)^6}
\sum_{\lambda_1\lambda_2}\Bigl\{
\bar\Psi_{\lambda_1\lambda_2\alpha'}(k_1,k_2,k^+_3)\,
[1+2\,\zeta{\cal P}_{12}]
\,j_{\alpha'\alpha}^\mu(k^+_3,k_3)\,
\Psi_{\lambda_1\lambda_2\alpha}(k_1,k_2,k_3)\nonumber\\
& &+\bar\Gamma_{\lambda_1\beta'\alpha}(k_1,k^+_2,k_3)
\,G_{\beta'\beta}(k_2^+)
\,j_{\beta\gamma}^\mu(k_2^+,k_2)\,
[1+2\,\zeta{\cal P}_{12}]\,u_{\gamma}(k_2,\lambda_2)\,
\Psi_{\lambda_1\lambda_2\alpha}(k_1,k_2,k_3)\nonumber\\
& &+\bar\Psi_{\lambda_1\lambda_2\alpha}(k_1,k_2,k^+_3)
\,\bar u_{\gamma}(k_2,\lambda_2)\,
[1+2\,\zeta{\cal P}_{12}]
\,j_{\gamma\beta'}^\mu(k_2,k^-_2)\,G_{\beta'\beta}(k_2^-)\,
\Gamma_{\lambda_1\beta\alpha}(k_1,k^-_2,k^+_3)
\Bigr\}\, , \label{CIA}
\end{eqnarray}
where the doubly off-shell vertex functions are evaluated
using (\ref{vertexoffshell}),
$j_{\alpha'\alpha}(k',k)$ is the single nucleon current for
off-shell nucleons with incoming (outgoing) four-momenta
$k^\mu$ ($k'^\mu$), $k_i^\pm=k_i\pm q$, and in every term
$k_1^2=k_2^2=m^2$ and $k_1+k_2+k_3=P$, where $P$ is the
four-momenta of the incoming deuteron.  Each diagram has two
off-shell propagators and three interactions, for a phase of
$i(-i)^5=1$.  The I diagrams are
\begin{eqnarray}
J^\mu_{\rm I}&=&3e\int\!\!\!\int\!\!\!\int
\frac{m^3\,d^3k_1d^3k_2d^3k'_2}{E_1E_2E'_2\,(2\pi)^9}
\sum_{\lambda_1\lambda_2\lambda_2'}
\bar\Psi_{\lambda_1\lambda'_2\alpha'}(k_1,k'_2,k'_3)\,
[1+2\,\zeta{\cal P}_{12}]
\nonumber\\
& &\qquad\qquad\qquad\qquad\qquad\qquad\qquad\times \;
V_{{\rm I}\,\lambda_2',\alpha';\lambda_2,\alpha}^\mu
(k'_2,k_3';k_2,k_3)\,[1+2\,\zeta{\cal P}_{12}]\,
\Psi_{\lambda_1\lambda_2\alpha}(k_1,k_2,k_3)\, ,
\label{IACelastic}
\end{eqnarray}
\end{widetext}
where $k_1+k_2+k_3=P$,
$k_1+k'_2+k'_3=P'=P+q$, $k_1^2=k'^2_2=k_2^2=m^2$,  and
$V_{{\rm I}\,\lambda_2',\alpha';\lambda_2,\alpha}^\mu
(k'_2,k_3';k_2,k_3)$ is the symmetrized two-body interaction
current for nucleons with incoming four-momenta $k_2,k_3$ and
outgoing four-momenta $k'_2,k'_3$. In $V^\mu_{\rm I}$ the first
four-momentum listed in each pair describes an on-shell
nucleon with incoming helicity $\lambda_2$ and outgoing
helicity $\lambda_2'$.  The second nucleon is off-shell, with
incoming Dirac index $\alpha$ and outgoing Dirac index
$\alpha'$.

The equations (\ref{CIA}) and (\ref{IACelastic}), with the
two particle off-shell vertex function defined by
Eq.~(\ref{vertexoffshell}), are convenient for numerical
calculations of the form factor at non-zero $q^2$.  However,
when $q\to0$, the nucleon propagators in the second and
third terms of the CIA result (\ref{CIA}) develop
singularities that cancel, leading to terms involving the
derivatives of the two-body kernel.  The Appendix
evaluates these diagrams in the $q\to0$ limit, and gives
an algebraic demonstration that the singularities cancel.  The
work is somewhat lengthy, and already contained implicitly
in the proof of gauge invariance.  A similar (but
algebraically different) demonstration of charge conservation
in the two-body case was already given in
Ref.~\cite{AVOG98}. 

We now turn to the expressions for the breakup current.

\subsection{The three-body breakup current}
\label{sec:D} 

The three-body breakup current was shown in Fig.~\ref{fig:breakup}.  The final
state in all the diagrams is antisymmetric.   As discussed in Sec.\ \ref{IIB}
this implies that the each diagram is multipied by  the projection operator
${\cal A}_3$ [defined in Eq.\ (\ref{A3})].  Hence the diagrams all have the
form given explicitly in Eq.\ (\ref{anti}).

The symmetrized RIA diagrams Fig.~\ref{fig:breakup} (a) are
\begin{widetext}
\begin{eqnarray}
J_{\rm RIA}^\mu&=&-{\cal A}_3\Bigg\{ \bar u_\gamma(p_3,\lambda_3) 
\Big[j^\mu_N(p_3, p_3^-)\Big]_{\gamma \gamma'}\,
\bar \Psi_{\lambda_1\lambda_2\gamma'}(p_1,p_2,p_3^-) 
+\zeta\;\bar u_\gamma(p_2,\lambda_2) 
\Big[j^\mu_N(p_2, p_2^-)\Big]_{\gamma \gamma'}\,
\bar \Psi_{\lambda_1\lambda_3\gamma'}(p_1,p_3,p_2^-) \Bigg\}
\nonumber\\
&=&-{\cal A}_3\Big(1 +\zeta {\cal P}_{23}\Big)\;\bar u_\gamma(p_3,\lambda_3) 
\Big[j^\mu_N(p_3, p_3^-)\Big]_{\gamma \gamma'}\,
\bar \Psi_{\lambda_1\lambda_2\gamma'}(p_1,p_2,p_3^-)\nonumber\\
&=&-2{\cal A}_3\,\bar u_\gamma(p_3,\lambda_3) 
\Big[j^\mu_N(p_3, p_3^-)\Big]_{\gamma \gamma'}\,
\bar \Psi_{\lambda_1\lambda_2\gamma'}(p_1,p_2,p_3^-)   \, . 
\label{RIAinelastic}
\end{eqnarray}  

Similarly, the interaction current diagrams, Fig.~\ref{fig:breakup} (c) and (d) are 
\begin{eqnarray}
J_{\rm I}^\mu&=&-{\cal A}_3\;\bar u_\gamma(p_3,\lambda_3) \,
\int\frac{m\,d^3k_2}{E_2\,(2\pi)^3}
V_{\rm I\,\lambda_2,\gamma;\lambda'_2,\gamma'}^\mu(p_2, p_3;k_2,k_3)\,
\Big[1+2\zeta{\cal P}_{12}\Big]\,
\bar \Psi_{\lambda_1\lambda'_2\gamma'}(p_1,k_2,k_3) 
\label{IACinelastic}
\end{eqnarray}  
and the final state interaction (FSI) terms, Fig.~\ref{fig:breakup} (b), arise from the core diagrams, and parallel those already given for the form factor in Eqs.\ (\ref{CIA}) and (\ref{IACelastic})
\begin{eqnarray}
J^\mu_{\rm FSI}&=&3e\int\!\!\int
\frac{m^2\,d^3k_1d^3k_2}{E_1E_2\,(2\pi)^6}
\sum_{\mu_1\mu_2}\nonumber\\
&&\Bigl\{
\bar T_{\lambda_1\lambda_2\lambda_3; \mu_1\mu_2\alpha''} (p_1,p_2,p_3;k_1,k_2,k^+_3) \,G_{\alpha''\alpha'}(k_3^+)\,
[1+2\,\zeta{\cal P}_{12}]
\,j_{\alpha'\alpha}^\mu(k^+_3,k_3)\,
\Psi_{\mu_1\mu_2\alpha}(k_1,k_2,k_3)\nonumber\\
& &+\bar T_{\lambda_1\lambda_2\lambda_3;\mu_1\beta'\alpha} (p_1,p_2,p_3;k_1,k^+_2,k_3)
\,G_{\beta'\beta}(k_2^+)
\,j_{\beta\gamma}^\mu(k_2^+,k_2)\,
[1+2\,\zeta{\cal P}_{12}]\,u_{\gamma}(k_2,\mu_2)\,
\Psi_{\mu_1\mu_2\alpha}(k_1,k_2,k_3)\nonumber\\
& &+\bar T_{\lambda_1\lambda_2\lambda_3;\mu_1\mu_2\alpha} (p_1,p_2,p_3;k_1,k_2,k^+_3)
\,\bar u_{\gamma}(k_2,\mu_2)\,
[1+2\,\zeta{\cal P}_{12}]
\,j_{\gamma\beta'}^\mu(k_2,k^-_2)\,G_{\beta'\beta}(k_2^-)\,
\Gamma_{\mu_1\beta\alpha}(k_1,k^-_2,k^+_3)
\Bigr\}\nonumber\\
&&+3e\int\!\!\!\int\!\!\!\int
\frac{m^3\,d^3k_1d^3k_2d^3k'_2}{E_1E_2E'_2\,(2\pi)^9}
\sum_{\lambda_1\lambda_2\lambda_2'}
\bar T_{\lambda_1\lambda_2\lambda_3;\mu_1\mu'_2\alpha''} (p_1,p_2,p_3;k_1,k'_2,k'_3)\,\, G_{\alpha''\alpha'}(k_3')
[1+2\,\zeta{\cal P}_{12}]
\nonumber\\
& &\qquad\qquad\qquad\qquad\qquad\qquad\qquad\times \;
V_{{\rm I}\,\mu_2',\alpha';\mu_2,\alpha}^\mu
(k'_2,k_3';k_2,k_3)\,[1+2\,\zeta{\cal P}_{12}]\,
\Psi_{\mu_1\mu_2\alpha}(k_1,k_2,k_3)
\, , \label{FSI}
\end{eqnarray}
where $T_{\lambda_1\lambda_2\lambda_3; \mu_1\mu_2\alpha} (p_1,p_2,p_3;k_1,k_2,k_3)$ is the symmetrized three-body scattering amplitude with the final state on-shell.  The off-shell unsymmetrized three-body scattering amplitude satisfies the equation shown in Fig.\ \ref{fig:scatteq}.  This is
\begin{eqnarray}
{\cal T}^{11}_{22\;\lambda_1\lambda_2\beta; \mu_1\mu_2\alpha} (p_1,p_2,p_3;k_1,k_2,k_3)
= \frac{1}{3}\frac{E_1}{m}\; (2\pi)^3\delta^3(p_1-k_1)\delta_{\lambda_1\mu_1} M^1_{22\;\lambda_2\beta,\mu_2\alpha}(p_2,k_2;P_{23})
\nonumber\\
-2\zeta\int\frac{m\,d^3k'_2}{E'_2\,(2\pi)^3}
M^1_{22\;\lambda_2\beta,\mu'_2\alpha'}(p_2,k'_2;P_{23})
G_{\alpha'\alpha''}(k'_3)
{\cal T}^{11}_{22\;\mu'_2\lambda_1\alpha''; \mu_1\mu_2\alpha} (k'_2,p_1,k'_3;k_1,k_2,k_3)\, .
\end{eqnarray}

The antisymmetrized amplitude is obtained by applying ${\cal A}_3$ to both the
initial and final state, and illustrated in Eq.\  (\ref{antit}).

\end{widetext}

\section{conclusions} 

Using diagrammatic techniques, we have derived a three-nucleon current
consistent with the three-body spectator equations, and have shown explicitly
that it is conserved.  We obtain the current for elastic scattering, shown in
Fig.\ \ref{fig:ff} and Eqs.\ (\ref{CIA}) and (\ref{IACelastic}), and for the
three-body breakup reaction, shown Fig.~\ref{fig:breakup} and Eqs.\
(\ref{RIAinelastic}), (\ref{IACinelastic}), and (\ref{FSI}).  The appendix also
shows explicitly that this current conserves the  charge of the bound state,
even with an arbitrary choice of electromagnetic form factors and in the
presence of energy dependent interactions.

Our results show that the spectator current will be free of any double counting
if we always choose to keep the spectator on-shell.  This simple rule leads to
an organizational principal that also resolves some ambiguities in the choice
of diagrams that might otherwise be present.

Our derivation and discussion relies heavily on the beautiful method of gauging
equations, developed by Kvinikhidze and Blankleider \cite{KB97,BK99}.   As
already noted in Sec.\ \ref{IIIB}, the normalization of the two-body and
three-body scattering amplitude used by KB is different from ours, giving
different weights to various diagrams, but the total results are the same.  

Another difference, also mentioned in  Sec.\ \ref{IIIB}, is that KB do not
rearrange their amplitudes so that the spectator is always on-shell.  This
rearrangement has some advantages; it not only makes the equations more
tractable, but it also displays how the spectator theory avoids the double
counting problem (recall Fig.\ \ref{fig:SpecDouble}), the importance of which
has been emphasized by Kvinikhidze and Blankleider.  

This rearrangement also leads to a different interpretation of the diagrams in
the theory.  For example, Eq.\ (60) of KB
includes RIA terms in which the photon couples to {\it all three\/} of the
final nucleons, while our RIA terms, given in Fig.\ \ref{fig:breakup} ($a_1$)
and ($a_2$), include only couplings to the particles in the final state
interacting pair, with {\it no\/} term describing coupling to the final state
spectator (both approaches require the results be antisymmetrized).

\begin{figure}[t]
\vspace*{0.4in}
\centerline{
\mbox{
\includegraphics[width=3in]{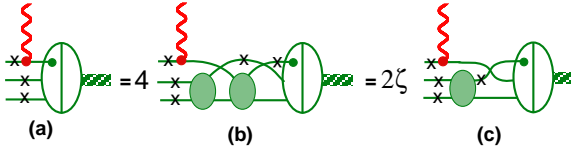}
}}
\caption{\footnotesize\baselineskip=10pt  (Color online) 
Diagram (a) is the coupling to the spectator particle included by KB as part of their RIA
contribution.  This diagram is missing from the RIA part of our result shown Fig.\ 
\ref{fig:breakup}(a), but can be transformed using the off-shell equations shown in Fig.\
\ref{fig:offshell} into diagram (c), included as part of the FSI shown in Fig.\
\ref{fig:breakup}(b).  The KB formalism does not include a term like (c), so both results are
identical.}
\vspace{0.2in}
\label{fig:RIA2}
\end{figure}  

At first glance it might seem that our result cannot be correct -- surely the
photon must couple to all of the outgoing nucleons.   But our total result {\it
does\/} include coupling to the final state spectator as part of the FSI term
discussed in Fig.\ \ref{fig:interpretfinal} (since this term arose from the
rearrangement, it is not present in KB).   In Fig.\ \ref{fig:RIA2}  the
off-shell equations of Fig.\ \ref{fig:offshell} are used to show
that the coupling to the final off-shell spectator given in KB [diagram (a)]
can be transformed into diagram (b) using the equation of Fig.\
\ref{fig:offshell}(c) which can be transformed using Fig.\
\ref{fig:offshell}(b) into the diagram of Fig.\  \ref{fig:interpretfinal}.  In
the KB approach, this term is part of the RIA, but in our approach it is part
of the FSI.  This shows that care must be taken in comparing the results of KB
with ours, but if this comparison is done carefully they appear to be in
agreement.

Independently, Adam and Van Orden \cite{AVO03} have derived the same current
operators using a completely algebraic approach along the lines developed in
Ref.\  \cite{AVOG98}.  They obtain the same results, with the same weight
factors.  Their result naturally leads to an ordering in which the spectator is
always on shell, in full agreement with us.  These two approaches reinforce and
complement each other. 

We plan to use the results of this paper to calculate the high energy breakup
processes recently measured at JLab.

\acknowledgements
   
It is a pleasure to acknowledge several helpful conversations with J.\ W.\ Van
Orden.  We also thank B. Blankleider and S. Kvinikhidze  for helpful
correspondence comparing our results with Ref.\ \cite{KB97}.  This work was
supported in part by the US Department of  Energy  under grant
No.~DE-FG02-97ER41032. The Southeastern Universities Research Association
(SURA) operates the Thomas Jefferson National Accelerator Facility under DOE
contract DE-AC05-84ER40150.  

M.T.P.\ and A.S.\ were supported in part by the Portuguese Funda\c c\~ao para a
Ci\^encia e a Tecnologia (FCT) under grant Nos.~CERN/FNU/43709/2001 and
POCTI/FNU/40834/2001. They thank the Theory Group of the Thomas Jefferson
National Accelerator Facility for the hospitality extended to them during their
recent visits.

\phantom{0}

\appendix


\section{Demonstration of charge conservation}

\begin{figure*}
\phantom{0}
\centerline{
\mbox{
\includegraphics[width=5.5in]{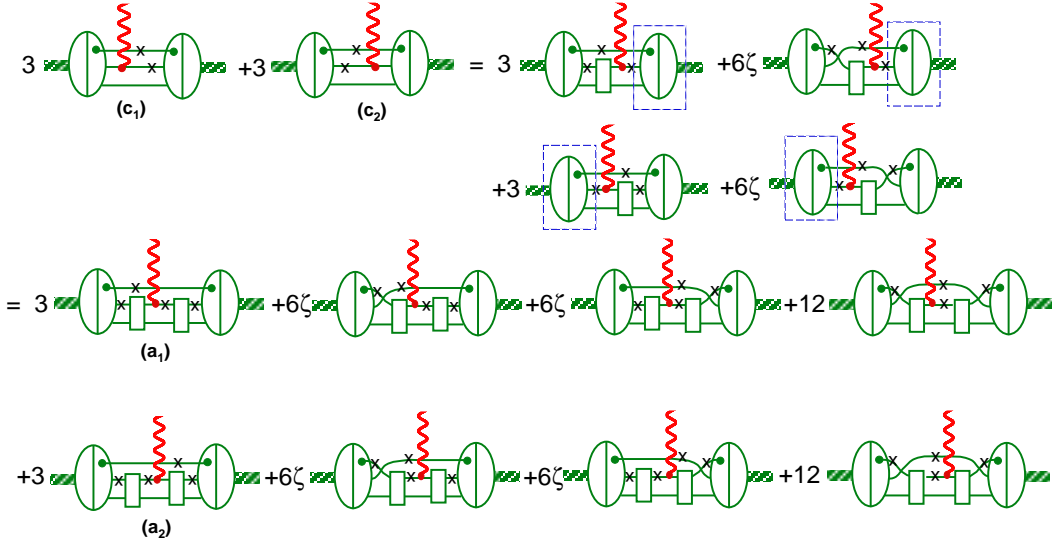}
}}
\caption{\footnotesize\baselineskip=10pt  (Color online) 
Diagrammatic representation of expansion of diagrams
\ref{fig:core} (c$_1$) and (c$_2$).  The first two rows use
Fig.~\ref{fig:offshell2}, and the second two rows use the
on-shell version of the same figure again (to
replace the amplitudes enclosed in the dashed boxes).  The
resulting 8 diagrams can be organized into 4 pairs, which
members of each pair arranged above and below each other on
lines 3 and 4 of the figure [diagrams (a$_1$) and (a$_2$) are
an example]. }
\label{fig:c1andc2}
\end{figure*}  

The explicit demonstration that charge is conserved begins by 
expanding the diagrams \ref{fig:core} (c$_1$) and (c$_2$)
using Fig.~\ref{fig:offshell2}.  The result is shown in
Fig.~\ref{fig:c1andc2}.  The 8 diagrams are organized into 4
pairs, with an integral over a three momentum ${\bf k}_2$.  All the
contributions from these 8 diagrams can be written  
\begin{widetext}
\begin{eqnarray}
J_{c1+c2}^\mu&=&3e\int\!\!\int\!\!
\int\frac{m^3\,d^3k_1d^3k''_2d^3k'_2}{E_1E''_2E'_2\,(2\pi)^9}\,
\bar\Psi_{\lambda_1\lambda'_2\alpha'} (k_1,k''_2,k''_3)\,
[1+2\zeta{\cal P}_{12}] 
\nonumber\\ & & \qquad\qquad\times
j^{\mu,\,c_1+c_2}_{\lambda'_2\alpha',\lambda_2\alpha}
(k_1,k''_2k''_3,k'_2k'_3)\,
[1+2\zeta{\cal P}_{12}]\,
\Psi_{\lambda_1\lambda_2\alpha}
(k_1,k'_2,k'_3)\,\, ,
\label{charge0}
\end{eqnarray}
where the integrand $j^\mu_c$ is the sum of two
contributions that each become singular as $q\to0$ 
\begin{eqnarray}
&&j^\mu_c\equiv
j^{\mu,\,c_1+c_2}_{\lambda'_2\alpha',\lambda_2\alpha}
(k_1,k''_2k''_3,k'_2k'_3)=
j^\mu_c
(k_1,k''_2P_{23}'',k'_2P_{23}')
\nonumber\\
&&=\int\frac{m\,d^3k_2}{E_2\,(2\pi)^3}\Bigg\{
V_{\lambda_2',\alpha';\beta',\gamma'}
(k''_2,k_2+q;P''_{23})
G_{\gamma'\gamma}(P'_{23}-k_2) 
\Big[\frac{1}{m-\not\!k_2-\not\!q}\,j^\mu_N(k_2+q,k_2)\,
\Lambda(k_2)\Big]_{\beta'\beta}\!\!
V_{\beta,\gamma;\lambda_2,\alpha}
(k_2,k'_2;P'_{23})
\nonumber\\
& &\qquad+V_{\lambda_2',\alpha';\beta',\gamma'}
(k''_2,k_2;P''_{23})
G_{\gamma'\gamma}(P''_{23}-k_2) 
\Big[\Lambda(k_2)\,j^\mu_N(k_2,k_2-q)
\,\frac{1}{m-\not\!k_2+\not\!q}
\Big]_{\beta'\beta}
V_{\beta,\gamma;\lambda_2,\alpha}
(k_2-q,k'_2;P'_{23}) \Bigg\}\, ,
\label{charge1}
\end{eqnarray}
%
%
with $j^\mu_N$ the single nucleon current and $\Lambda(k_2)=(m+\not\!k_2)/(2m)$ the positive energy projection operator for a particle with four-momentum $k_2$.
The integrals in (\ref{charge1}) are both given in terms of
the on-shell four momentum $k_2=\{E_2,{\bf k}_2\}$.  The
two-body  kernels, $V$, conserve four-momentum, and are
expressed in terms of independent variables, which we choose
to be the initial and final momenta of particle 2
(usually on-shell) and the total momentum of the two-body
system, so that generically
\begin{eqnarray}
V\equiv V(k_2',k_3';k_2,k_3)= V(k'_2,k_2;P_{23})
\end{eqnarray}
with $P_{23}=P-k_1=k_2+k_3=k'_2+k'_3$.

To calculate the $q\to0$ limit of (\ref{charge1}), we follow
the method developed in Ref.~\cite{AGS97}.   Since the full
current is gauge invariant, we know that the singular terms
must vanish as $q\to0$, so we can find the $q\to0$
limit of (\ref{charge1}) by contracting $q_\mu$ into both
sides, expanding about $q_\mu=0$, and extracting the
coefficient of the linear term.  Using the Ward identity for
the nucleon current
\bea
q_\mu \,j^\mu_N(k',k)= (m-\not\!\!k)-(m-\not\!k')
\eea 
and, for $k_2$ on-shell,
\bea
(m-\not\!k_2)\Lambda(k_2)=0\, ,
\eea
Eq.~(\ref{charge1}) reduces to
%
\bea
q_\mu\,j^\mu_c&=&
-\int\frac{m\,d^3k_2}{E_2\,(2\pi)^3}
\Bigg\{
V_{\lambda_2',\alpha';\beta',\gamma'}
(k''_2,k_2+q;P''_{23})
G_{\gamma'\gamma}(P'_{23}-k_2) \,
\Lambda(k_2)_{\beta'\beta}
V_{\beta,\gamma;\lambda_2,\alpha}
(k_2,k'_2;P'_{23})
\nonumber\\
& &\qquad\qquad\qquad-V_{\lambda_2',\alpha';\beta',\gamma'}
(k''_2,k_2;P''_{23})
G_{\gamma'\gamma}(P''_{23}-k_2) 
\,\Lambda(k_2)_{\beta'\beta}
V_{\beta,\gamma;\lambda_2,\alpha}
(k_2-q,k'_2;P'_{23}) \Bigg\}
\nonumber\\
&=&-q_\mu\int\frac{m\,d^3k_2}{E_2\,(2\pi)^3} 
\sum_{\lambda''_2}\Bigg\{\left[\frac{\partial}{\partial
k_{2\,\mu}}V_{\lambda_2',\alpha';\lambda''_2,\gamma'}
(k''_2,k_2;P'_{23})\right] G_{\gamma'\gamma}(P'_{23}-k_2) \,
V_{\lambda_2'',\gamma;\lambda_2,\alpha}
(k_2,k'_2;P'_{23}) \nonumber\\
&&\qquad\qquad\qquad+V_{\lambda_2',\alpha';\lambda''_2,\gamma'}
(k''_2,k_2;P'_{23}) G_{\gamma'\gamma}(P'_{23}-k_2) \,
\left[\frac{\partial}{\partial
k_{2\,\mu}} V_{\lambda_2'',\gamma;\lambda_2,\alpha}
(k_2,k'_2;P'_{23}) \right] \nonumber\\
&&\qquad\qquad\qquad-V_{\lambda_2',\alpha';\lambda''_2,\gamma'}
(k''_2,k_2;P'_{23}) 
\left[\frac{\partial}{\partial
P'_{23\,\mu}} G_{\gamma'\gamma}(P'_{23}-k_2) \right] \,
V_{\lambda_2'',\gamma;\lambda_2,\alpha}
(k_2,k'_2;P'_{23}) \Bigg\}+ {\cal O}(q^2)\, .\quad
\label{charge2}
\eea
Use the shorthand notation
\begin{eqnarray}
\frac{\partial}{\partial k_{2\,\mu}}V(k''_2,k_2;P'_{23})
\equiv \delta^\mu_i V\, ;\quad
\frac{\partial}{\partial k_{2\,\mu}}V(k_2,k'_2;P'_{23})\equiv
\delta^\mu_f  V\, ;\quad \frac{\partial G(k_3)}{\partial
k_{3\,\mu}}
\equiv G^\mu(k_3)
\label{delta}
\end{eqnarray}
where, in general, $\delta^\mu_i$ refers to the derivative
with respect to the initial $k_2$ momentum, and $\delta^\mu_f$
with respect to the final $k_2$ momentum.  With this
notation  
\begin{eqnarray}
\lim_{q\to0}j^\mu_c&=&
\int\frac{m\,d^3k_2}{E_2\,(2\pi)^3}\sum_{\lambda''_2}\;
\Big\{V_{\lambda_2',\alpha';\lambda''_2,\gamma'}
(k''_2,k_2;P'_{23})\;
G^{\mu}_{\gamma'\gamma}(k_3)\,
V_{\lambda''_2,\gamma;\lambda_2,\alpha}
(k_2,k'_2;P'_{23})
\nonumber\\
&-&
\delta^\mu_i V_{\lambda_2',\alpha';\lambda''_2,\gamma'}
(k''_2,k_2;P'_{23})
\,G_{\gamma'\gamma}(k_3)\,
V_{\lambda''_2,\gamma;\lambda_2,\alpha}
(k_2,k'_2;P'_{23})
\nonumber\\
&-& V_{\lambda_2',\alpha';\lambda''_2,\gamma'}
(k''_2,k_2;P'_{23})
\,G_{\gamma'\gamma}(k_3)\;\delta^\mu_f
V_{\lambda''_2,\gamma;\lambda_2,\alpha}
(k_2,k'_2;P'_{23})\Big\}\, .
\label{charge3}
\end{eqnarray}
Inserting this back into the original expression
(\ref{charge0}), and using the wave equation
[(\ref{vertexoffshell}) with the second particle on-shell],
gives
\begin{eqnarray}
\lim_{q\to0}J^\mu_{c_1+c_2}=
3e\int\!\!\int\frac{m^2\,d^3k_1d^3k_2}{E_1E_2\,(2\pi)^6}
\sum_{\lambda_1\lambda_2}\;
\Bigg\{\bar\Gamma_{\lambda_1\lambda_2\gamma'}
(k_1,k_2,k_3)\;
G^{\mu}_{\gamma'\gamma}(k_3)\,
\Gamma_{\lambda_1\lambda_2\gamma}
(k_1,k_2,k_3)\qquad\qquad\qquad\qquad\qquad
\nonumber\\
+\int\frac{m\,d^3k'_2}{E'_2\,(2\pi)^3}\sum_{\lambda'_2}
\bar\Psi_{\lambda_1\lambda'_2\alpha'}
(k_1,k'_2,k'_3)\,[1+2\zeta{\cal P}_{12}]\,
\delta^\mu_i V_{\lambda_2',\alpha';\lambda_2,\alpha}
(k'_2,k_2;P_{23})
\,\Psi_{\lambda_1\lambda_2\alpha}
(k_1,k_2,k_3)\qquad\quad
\nonumber\\
+ \int\frac{m\,d^3k'_2}{E'_2\,(2\pi)^3}\sum_{\lambda'_2}
\bar\Psi_{\lambda_1\lambda'_2\alpha'} (k_1,k'_2,k'_3)\;
\delta^\mu_f V_{\lambda'_2,\alpha';\lambda_2,\alpha}
(k'_2,k_2;P_{23})\,[1+2\zeta{\cal
P}_{12}]\,\Psi_{\lambda_1\lambda_2\alpha}
(k_1,k_2,k_3)\Bigg\}\, .\quad\;\;
\label{charge4}
\end{eqnarray}
Note that the relative signs of the terms change because of
the sign in Eq.~(\ref{vertexoffshell}).

\begin{figure*}
\vspace*{-0.2in}
\centerline{
\mbox{
\includegraphics[width=5in]{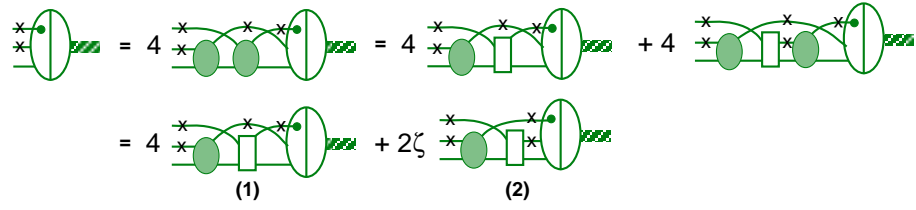}
}}
\caption{\footnotesize\baselineskip=10pt  (Color online) 
The second two-body scattering amplitude in
Fig.~\ref{fig:offshell} (c) can be replaced by its integral
equation, and further simplified using the on-shell version
of \ref{fig:offshell} (b).  This form of the vertex function
that is useful in the reduction of diagrams
\ref{fig:core} (e$_1$) and (e$_2$). }
\label{fig:offshell3}
\end{figure*}  
\begin{figure*}
%
\centerline{
\mbox{
\includegraphics[width=6.4in]{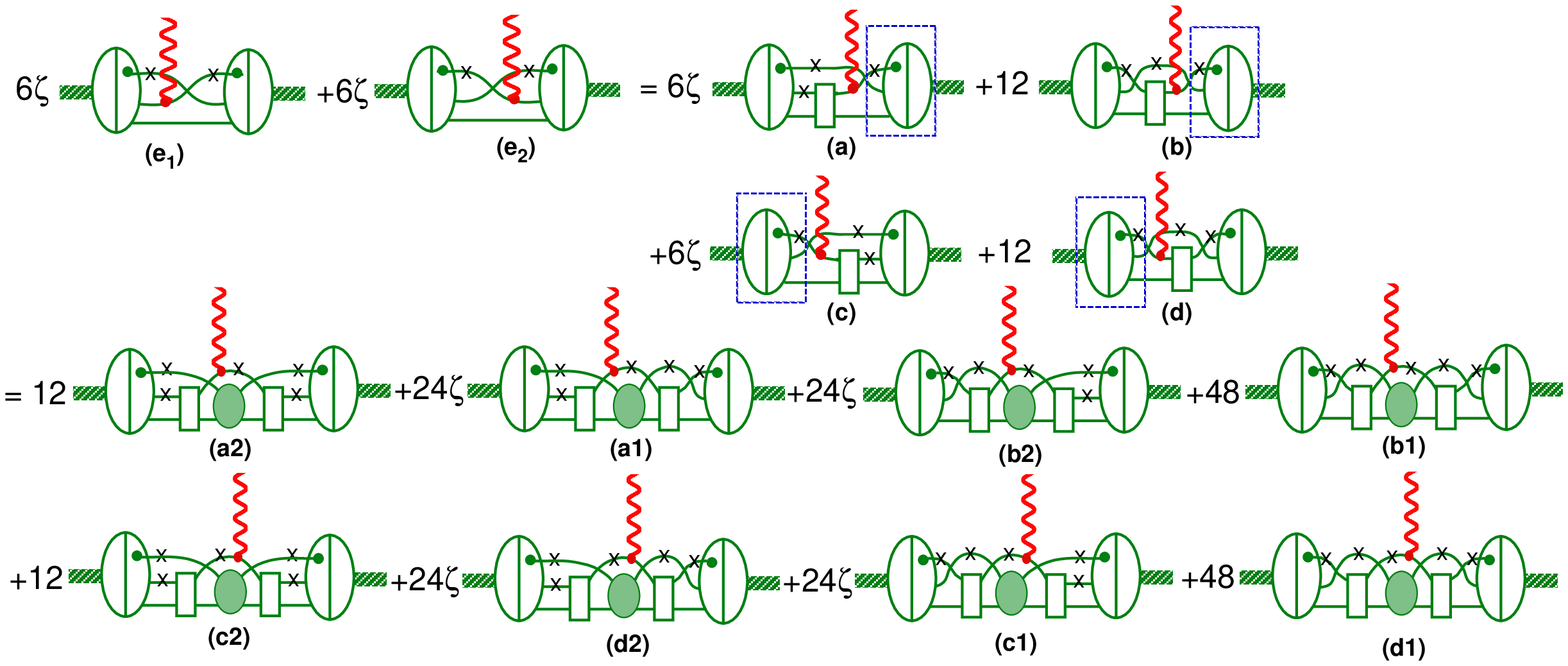}
}}
\caption{\footnotesize\baselineskip=10pt  (Color online) 
Diagrammatic representation of expansion of diagrams
\ref{fig:core} (e$_1$) and (e$_2$).  The first two rows use
Fig.~\ref{fig:offshell2}, and the second two rows use
Fig.~\ref{fig:offshell3} (to replace the amplitudes enclosed
in the dashed boxes).  The resulting 8 diagrams are labeled
by both the parent diagram (a--d in the first two rows) and
the term (1 or 2) in Fig.~\ref{fig:offshell3} from which they
originate.  They are organized into 4 pairs, with members
of each pair arranged above and below each other on lines 3
and 4 of the figure [diagrams (a2) and (c2) are an
example].}
\label{fig:e1ande2}
\end{figure*}  

We now turn to diagrams Fig.~\ref{fig:core} (e$_1$) 
and (e$_2$).  The object here is to isolate the
electromagnetic coupling in a loop as we did above, but
because the loop now involves a spectator, we need to pull out
{\it two\/} interactions.  In preparation, we note that the
Fig.~\ref{fig:offshell} (c) can be written in an alternative
form that involves the kernel, as shown in
Fig.~\ref{fig:offshell3}.  The algebraic form of this
equation is  
\begin{eqnarray}
\Gamma_{\lambda_1\lambda_2\alpha}(k_1,k_2,k_3)&=&
2\zeta\,{\cal P}_{12}\int
\frac{m^2\,d^3k'_1d^3k'_2}{E'_1E_2'\,(2\pi)^6}
\sum_{\lambda_1'\lambda_2'}
M_{\lambda_1\alpha,\lambda'_1\gamma'}(k_1,k'_1;P_{13})\,
G_{\gamma'\gamma}(P_{13}-k'_1)\nonumber\\
& &\qquad\qquad\times
V_{\lambda_2\gamma,\lambda_2'\alpha'}(k_2,k_2';P'_{23})\,
[1+2\,\zeta\,{\cal P}_{12}]\,
\Psi_{\lambda'_1\lambda'_2\alpha'}(k'_1,k'_2,k'_3)
\, ,
\label{pullout2}
\end{eqnarray}
where $P_{13}=P-k_2$ and $P'_{23}=P-k'_1$.

Using this substitution, the diagrams
Fig.~\ref{fig:core} (e$_1$) and (e$_2$) can be written as
shown in Fig.~\ref{fig:e1ande2}.  These 8 diagrams collect
together into  
\begin{eqnarray}
J_{e1+e2}^\mu&=&-12e\int\!\!\int\!\!\int\!\!
\int\frac{m^4\,d^3k''_1d^3k'_1d^3k''_2d^3k'_2}{E''_1E'_1E''_2E'_2\,(2\pi)^{12}}\,
\bar\Psi_{\lambda'_1\lambda'_2\alpha'} (k''_1,k''_2,k''_3)\,
[1+2\zeta{\cal P}_{12}] 
\nonumber\\ & & \qquad\qquad\qquad\times
j^{\mu,\,e_1+e_2}_{\lambda'_1\lambda'_2\alpha',
\lambda_1\lambda_2\alpha}
(k''_1k''_2k''_3,k'_1k'_2k'_3)\,
[1+2\zeta{\cal P}_{12}]\,
\Psi_{\lambda_1\lambda_2\alpha}
(k'_1,k'_2,k'_3)\,\, ,
\label{charge0a}
\end{eqnarray}
where the common internal loop is again the sum of two
(canceling) singular terms
\begin{eqnarray}
j^\mu_e\equiv
j^{\mu,\,e_1+e_2}_{\lambda'_1\lambda'_2\alpha',
\lambda_1\lambda_2\alpha}(k''_1k''_2k''_3,k'_1k'_2k'_3)
=j^\mu_e(k''_1k''_2,P''_{23};k'_1k'_2,P'_{23})
=\int\frac{m\,d^3k_2}{E_2\,(2\pi)^3}\qquad\qquad\qquad\qquad
\qquad\qquad\qquad\qquad
\nonumber\\
\Bigg\{V_{\lambda_2',\alpha';\beta',\gamma'}
(k''_2,k_2+q;P''_{23})\!\!
\left[\frac{1}{m-\not\!k_2-\not\!q}\;j_N^\mu(k_2+q,k_2)\;
\Lambda(k_2)
\right]_{\beta'\beta}\!\!V_{\beta,\gamma;\lambda_2,\alpha}
(k_2,k'_2;P'_{23})
\,{\cal O}_{\lambda'_1\gamma',\lambda_1\gamma}(k''_1,k'_1;P_{13}-q)
\quad
\nonumber\\
+V_{\lambda_2',\alpha';\beta',\gamma'}
(k''_2,k_2;P''_{23})\!\!
\left[\Lambda(k_2)\;j_N^\mu(k_2,k_2-q)\;
\frac{1}{m-\not\!k_2+\not\!q}
\right]_{\beta'\beta}\!\!V_{\beta,\gamma;\lambda_2,\alpha}
(k_2-q,k'_2,P'_{23})
\,{\cal O}_{\lambda'_1\gamma',\lambda_1\gamma}(k''_1,k'_1;P_{13}) 
\Bigg\} \, ,\qquad
\label{charge1a}
\end{eqnarray}
where
\begin{eqnarray}
{\cal
O}_{\lambda'_1\gamma',\lambda_1\gamma}
(k''_1,k'_1;P_{13})
=G_{\gamma'\omega'}(P_{13}-k''_1)\,
M_{\lambda'_1\omega',\lambda_1\omega}(k_1'',k'_1;P_{13})\,
G_{\omega\gamma}(P_{13}-k'_1)
\end{eqnarray}
and $P'_{23}=P-k'_1$, $P''_{23}=P'-k''_1$, $P'=P+q$, and 
$P_{13}=P'-k_2$.  Following the
same arguments that lead from Eq.~(\ref{charge1}) to
(\ref{charge3}), we obtain
\begin{eqnarray}
\lim_{q\to0}j^\mu_e&=&
\int\frac{m\,d^3k_2}{E_2\,(2\pi)^3}\sum_{\lambda''_2}\;
\Big\{V_{\lambda_2',\alpha';\lambda''_2,\gamma'}
(k''_2,k_2;P''_{23})\,
\Delta^\mu{\cal O}_{\lambda'_1\gamma',\lambda_1\gamma}
(k''_1,k'_1;P_{13})\,
V_{\lambda''_2,\gamma;\lambda_2,\alpha} (k_2,k'_2;P'_{23})
\nonumber\\
&&\qquad\qquad\qquad\qquad-\,
\delta^\mu_i V_{\lambda_2',\alpha';\lambda''_2,\gamma'}
(k''_2,k_2;P_{23}'')
\,{\cal
O}_{\lambda'_1\gamma',\lambda_1\gamma} 
(k''_1,k'_1;P_{13})\,
V_{\lambda''_2,\gamma;\lambda_2,\alpha}
(k_2,k'_2;P_{23}')
\nonumber\\
&&\qquad\qquad\qquad\qquad-\, 
V_{\lambda_2',\alpha';\lambda''_2,\gamma'}
(k''_2,k_2;P_{23}'')
\,{\cal
O}_{\lambda'_1\gamma',\lambda_1\gamma}
(k''_1,k'_1;P_{13})
\;\delta^\mu_f V_{\lambda''_2,\gamma;\lambda_2,\alpha}
(k_2,k'_2;P_{23}')\Big\}\, ,
\label{charge3a}
\end{eqnarray}
where $\delta^\mu_i V$ and $\delta^\mu_fV$ have been
previously defined, and
\begin{eqnarray}
\Delta^\mu{\cal O}_{\lambda'_1\gamma',\lambda_1\gamma}
(k''_1,k'_1;P_{13})
&=&G^{\mu}_{\gamma'\omega'}(P_{13}-k''_1)\;
M_{\lambda'_1\omega',\lambda_1\omega}(k_1'',k'_1;P_{13})\,
G_{\omega\gamma}(P_{13}-k'_1) \nonumber\\
&+& 
G_{\gamma'\omega'}(P_{13}-k''_1)\;
M_{\lambda'_1\omega',\lambda_1\omega}
(k_1'',k'_1;P_{13})\,
G^{\mu}_{\omega\gamma}(P_{13}-k'_1)
\nonumber\\
&+& 
G_{\gamma'\omega'}(P_{13}-k''_1)\;
\delta^\mu_P M_{\lambda'_1\omega',\lambda_1\omega}
(k_1'',k'_1;P_{13})\,
G_{\omega\gamma}(P_{13}-k'_1)
\, ,
\label{oanddo}
\end{eqnarray}
with
\bea
\delta^\mu_P M(k_1'',k'_1;P_{13})\equiv 
\left(\frac{\partial }{\partial P_{13}} \right)^\mu M(k_1'',k'_1;P_{13})
\eea
the  derivative with respect to the total two-body momentum. 
Inserting (\ref{charge3a}) into (\ref{charge0a}), and using
the replacement (\ref{oanddo}) and the wave equations
(\ref{vertexoffshell}) and (\ref{pullout2}), gives an
intermediate result  
\begin{eqnarray}
&&\!\!\!\!\!\!\!\!\lim_{q\to0}J^\mu_{e_1+e_2}=
3 e\int\!\!\int\frac{m^2\,d^3k_1d^3k_2}{E_1E_2\,(2\pi)^6} \sum_{\lambda_1\lambda_2}\;
\Bigg\{\bar\Gamma_{\lambda_1\lambda_2\gamma'}
(k_1,k_2,k_3)\;
G^{\mu}_{\gamma'\gamma}(k_3)\,4\zeta{\cal P}_{12}\,
\Gamma_{\lambda_1\lambda_2\gamma}
(k_1,k_2,k_3)\qquad\qquad\qquad\qquad
\nonumber\\ 
&&\qquad\qquad\qquad\qquad+\int\frac{m\,d^3k'_2}{E'_2\,(2\pi)^3}
\sum_{\lambda'_2}
\bar\Psi_{\lambda_1\lambda'_2\alpha'}
(k_1,k'_2,k'_3)\, 
{\cal X}^\mu_{\lambda_2',\alpha';\lambda_2,\alpha}
(k'_2,k_2;P_{23})\,\Psi_{\lambda_1\lambda_2\alpha}
(k_1,k_2,k_3)
\Bigg\}\, .\qquad
\label{charge2a}
\end{eqnarray}
where
\begin{eqnarray}
{\cal X}^\mu_{\lambda_2',\alpha';\lambda_2,\alpha}
(k'_2,k_2;P_{23})&=&[1+2\zeta{\cal P}_{12}]\,
\delta^\mu_i
V_{\lambda_2',\alpha';\lambda_2,\alpha}
(k'_2,k_2;P_{23})
\,2\zeta{\cal P}_{12}
+2\zeta{\cal P}_{12}\;
\delta^\mu_f
V_{\lambda'_2,\alpha';\lambda_2,\alpha}
(k'_2,k_2;P_{23})\,[1+2\zeta{\cal
P}_{12}]
\nonumber\\
&&-2\zeta{\cal P}_{12}\,
\delta^\mu_PM_{\lambda'_2,\alpha';\lambda_2,\alpha}
(k'_2,k_2;P_{32})\;2\zeta{\cal
P}_{12}\, .
\label{chidef}
\end{eqnarray}

The evaluation of the derivative of $M$ can be carried out
using the fact that $M$ is an infinite series of interactions:
\begin{eqnarray}
M=V-VGV+VGVGV-\cdots\, ,
\label{Mseries}
\end{eqnarray}
and recalling that $k_3=P_{23}-k_2$, so that $\partial
G(k_3)/\partial (P_{23})_\mu=\partial
G(k_3)/\partial (k_3)_\mu=G^\mu$   
\begin{eqnarray}
\delta^\mu_PM
&=&\delta^\mu_PV-(\delta^\mu_PV) GV-VG^\mu V-
VG(\delta^\mu_PV)+(\delta^\mu_PV) GVGV+VG^\mu VGV+VG(\delta^\mu_PV)GV
+\cdots\nonumber\\  
&=&\delta^\mu_PV-(\delta^\mu_PV) GM-MG(\delta^\mu_PV) 
-MG^\mu M+MG(\delta^\mu_PV) GM\, .
\label{Mderivative}
\end{eqnarray}
Therefore, the $\delta^\mu_PM$ term in
(\ref{chidef}) becomes 
\begin{eqnarray}
\delta^\mu_PM
&=& 
\delta^\mu_PV_{\lambda'_2\alpha',\lambda_2\alpha}(k_2',k_2;P_{23})
-\int\frac{m\,d^3k''_2}{E''_2\,(2\pi)^3}\sum_{\lambda''_2}
\delta^\mu_PV_{\lambda'_2\alpha',\lambda''_2\beta'}
(k_2',k''_2;P_{23})\,G_{\beta'\beta}( k''_3)
M_{\lambda''_2\beta,\lambda_2\alpha}
(k_2'',k_2;P_{23})
\nonumber\\
&&-\int\frac{m\,d^3k''_2}{E''_2\,(2\pi)^3}\sum_{\lambda''_2}
M_{\lambda'_2\alpha',\lambda''_2\beta'}
(k_2',k''_2;P_{23})\,G_{\beta'\beta}( k''_3)
\delta^\mu_PV_{\lambda''_2\beta,\lambda_2\alpha}
(k_2'',k_2;P_{23})
\nonumber\\
&&-\int\frac{m\,d^3k''_2}{E''_2\,(2\pi)^3}\sum_{\lambda''_2}
M_{\lambda'_2\alpha',\lambda''_2\beta'}
(k_2',k''_2;P_{23})\,
G^\mu_{\beta'\beta}( k''_3)
M_{\lambda''_2\beta,\lambda_2\alpha}
(k_2'',k_2;P_{23})
\nonumber\\
&&+\int\frac{m^2\,d^3k''_2\,d^3k'''_2}{E''_2E'''_2\,(2\pi)^6}
\sum_{\lambda''_2\lambda'''_2} 
M_{\lambda'_2\alpha',\lambda''_2\beta'}
(k_2',k''_2;P_{23})\,G_{\beta'\beta}( k''_3)
\delta^\mu_PV_{\lambda''_2\beta,\lambda'''_2\rho'}
(k_2'',k'''_2;P_{23})
\nonumber\\
&&\qquad\qquad\qquad\qquad\times G_{\rho'\rho}( k'''_3)\,
M_{\lambda'''_2\rho,\lambda_2\alpha}
(k_2''',k_2;P_{23})
\, .
\label{Mreduce}
\end{eqnarray}
Using this expression and the original form of the wave
equation, (\ref{boundeq}),  the term involving the
derivative of $M$ reduces to   
\begin{eqnarray}
&&\!\!\!\!\!\!\!\!\lim_{q\to0}
J^\mu_{e_1+e_2}\Big|_{M^\mu\,{\rm term}}= 3
e\int\!\!\int\frac{m^2\,d^3k_1d^3k_2}{E_1E_2\,(2\pi)^6}
\sum_{\lambda_1\lambda_2}\;
\Bigg\{\bar\Gamma_{\lambda_1\lambda_2\gamma'}
(k_1,k_2,k_3)\;
G^{\mu}_{\gamma'\gamma}(k_3)\,
\Gamma_{\lambda_1\lambda_2\gamma}
(k_1,k_2,k_3)
\nonumber\\
&-&\int\frac{m\,d^3k'_2}{E'_2\,(2\pi)^3}\sum_{\lambda'_2}
\bar\Psi_{\lambda_1\lambda'_2\alpha'}
(k_1,k'_2,k'_3)\,
[1+2\zeta{\cal P}_{12}]\,
\delta^\mu_PV_{\lambda'_2\alpha',\lambda_2\alpha}
(k_2',k_2;P_{23})  
\,[1+2\zeta{\cal P}_{12}]\,
\Psi_{\lambda_1\lambda_2\alpha} (k_1,k_2,k_3)
\Bigg\}\, .\qquad
\label{charge4a}
\end{eqnarray}
Collecting all
terms from Eqs.~(\ref{charge4}), (\ref{charge2a}), and
(\ref{charge4a}) gives   
\begin{eqnarray}
\lim_{q\to0}J^\mu_{c+e}=
3 e\int\!\!\int\frac{m^2\,d^3k_1d^3k_2}{E_1E_2\,(2\pi)^6}
\sum_{\lambda_1\lambda_2}\;
\Bigg\{\bar\Gamma_{\lambda_1\lambda_2\gamma'}
(k_1,k_2,k_3)\;
G^{\mu}_{\gamma'\gamma}(k_3)\,[2+4\zeta{\cal P}_{12}]\,
\Gamma_{\lambda_1\lambda_2\gamma}
(k_1,k_2,k_3)
\nonumber\\
+\int\frac{m\,d^3k'_2}{E'_2\,(2\pi)^3}\sum_{\lambda'_2}
\bar\Psi_{\lambda_1\lambda'_2\alpha'}
(k_1,k'_2,k'_3)\,[1+2\zeta{\cal P}_{12}]\,(\Delta V)^\mu_{c+e}\,
[1+2\zeta{\cal
P}_{12}]\,
\Psi_{\lambda_1\lambda_2\alpha}
(k_1,k_2,k_3)
\Bigg\}\, .
\label{charge5}
\end{eqnarray}
where
\begin{eqnarray}
(\Delta V)^\mu_{c+e} =
\Big(\delta^\mu_i +\delta^\mu_f-\delta^\mu_P
\Big)
V_{\lambda_2',\alpha';\lambda_2,\alpha} (k'_2,k_2;P_{23})
\, .
\label{vdelta1}
\end{eqnarray}

We next look at the $q\to0$ limit of the interaction current
terms, shown in Figs.~\ref{fig:core} (f) and written in
Eq.~(\ref{IACelastic}).  Rewriting the WT identity for the
interaction current, Eq.~(\ref{WTonV}), in terms of the independent variables gives
\bea
q_\mu V^\mu_{\rm I}=
V(k'_2,k_2;P_{23})- V(k'_2,k_2;P_{23}+q)
+V(k'^-_2,k_2;P_{23})- V(k'_2,k^+_2;P_{23}+q)
\, .
\label{WTonV2}
\eea
Expanding (\ref{WTonV2})  in powers of
$q^\mu$, and equating the coefficient of the term linear in
$q^\mu$, gives 
\begin{eqnarray}
\lim_{q\to0}V^\mu_{{\rm I}\,\lambda'_2,\alpha';\lambda_2;\alpha}
(k'_2,k_2;P_{23})=- \Big\{\delta_i^\mu +\delta_f^\mu
+2\delta_P^\mu \Big\} V_{\lambda'_2,\alpha';\lambda_2;\alpha}
(k'_2,k_2;P_{23})
\, .
\label{IACat0}
\end{eqnarray}
Substituting this into (\ref{IACelastic}) gives a result with
the same form as the second term in (\ref{charge5}), and combining this with (\ref{vdelta1}) gives a total contribution of $-3\delta_P^\mu V$.  Combining this with the contributions from diagrams \ref{fig:core} (b) and (d), and the other terms from (\ref{charge5}), gives the total result
\begin{eqnarray}
J^\mu_{\rm total}(0)&=& 9 e\int\!\!\int\frac{m^2\,d^3k_1d^3k_2}{E_1E_2\,(2\pi)^6}
\sum_{\lambda_1\lambda_2}\;
\bar\Gamma_{\lambda_1\lambda_2\gamma'}
(k_1,k_2,k_3)\;
G^{\mu}_{\gamma'\gamma}(k_3)\,[1+2\zeta{\cal P}_{12}]\,
\Gamma_{\lambda_1\lambda_2\gamma}
(k_1,k_2,k_3)\nonumber\\
&& -9e
\int\!\!\int\!\!\int\frac{m^3\,d^3k_1d^3k_2d^3k'_2}{E_1E_2E'_2\,(2\pi)^9}
\sum_{\lambda_1\lambda_2\lambda'_2}
\bar\Psi_{\lambda_1\lambda'_2\alpha'}
(k_1,k'_2,k'_3)
[1+2\zeta{\cal P}_{12}] (\delta^\mu_P V)  [1+2\zeta{\cal P}_{12}]\,
\Psi_{\lambda_1\lambda_2\alpha}
(k_1,k_2,k_3)\qquad
\nonumber\\
&=&\lim_{q\to0} \,3e\,\bar u(P',\lambda')\left[ F_1(Q^2)\gamma^\mu +\frac{i\sigma^{\mu\nu} q_\nu }{2M_B} F_2(Q^2) \right] u(P,\lambda)
=3e\,\frac{P^\mu}{M_B}\, ,
\end{eqnarray}  
where the last line uses the fact that the total charge of a bound state of three identical particles of charge $e$ is $3e$ (because isospin has been ignored).   Hence charge is conserved if the normalization of the wave function is
\begin{eqnarray}
1= 3\int\!\!\int\left<
\bar\Gamma\;
\frac{P^\mu}{M_B} \frac{\partial G}{\partial P^\mu} 
\,[1+2\zeta{\cal P}_{12}]\,
\Gamma\right>
 -3\int\!\!\int\!\!\int\left<\bar \Psi\; [1+2\zeta{\cal P}_{12}]\,
 \frac{P_\mu}{M_B}\frac{\partial V}{\partial P^\mu} \,
  [1+2\zeta{\cal P}_{12}]  \; \Psi \right> \, . \qquad\;
\end{eqnarray}  
Noting that 
\bea
\frac{P_\mu}{M_B}\frac{\partial G}{\partial P^\mu}= 2M_B\frac{\partial G}{\partial P^2} \, ,
\eea
and similarly for $V$, we recover a normalization condition equal to $2M_B$ times that originally derived in Ref.\ \cite{AGS97}.  This difference is due to the fact that the all spinors in Ref.\ \cite{AGS97} the were normalized to $2m$ (or $2M_B$ for the bound state), so our result agrees with Ref.\ \cite{AGS97}, completing our demonstration.

\phantom{0}

\end{widetext}

\end{document}